\documentclass[graybox, envcountchap]{svmult}

\usepackage{makeidx}         % allows index generation
\usepackage{graphicx}        % standard LaTeX graphics tool
\usepackage{multicol}        % used for the two-column index
\usepackage[bottom]{footmisc}% places footnotes at page bottom

\usepackage{newtxtext}       % 
\usepackage{newtxmath}       % selects Times Roman as basic font
\UseRawInputEncoding
\usepackage{comment} 
\usepackage{amsmath}
\usepackage{tikz}
\usepackage{hyperref}
\usepackage[utf8]{inputenc}
\usepackage[T1]{fontenc}

\DeclareMathOperator{\arccosh}{arccosh}
\newcommand{\be}{\begin{equation}}
\newcommand{\ee}{\end{equation}}
\newcommand{\beq}{\begin{eqnarray}}
\newcommand{\eeq}{\end{eqnarray}}
\newcommand{\ba}{\begin{align}}
\newcommand{\ea}{\end{align}}

\makeindex             % used for the subject index
\begin{document}

\title*{Deformed relativistic symmetry principles}
\author{Michele Arzano, Giulia Gubitosi and Jos\'e Javier Relancio}
\institute{Michele Arzano \at Dipartimento di Fisica ``Ettore Pancini'', Universit\`{a} di Napoli {\sl Federico II}\vspace{5pt}, 80125 Naples, Italy, and INFN, Sezione di Napoli, 80125 Naples, Italy, \email{michele.arzano@na.infn.it}
\and  Giulia Gubitosi \at Dipartimento di Fisica ``Ettore Pancini'', Universit\`{a} di Napoli {\sl Federico II}\vspace{5pt}, 80125 Naples, Italy and INFN, Sezione di Napoli, 80125 Naples, Italy,\email{giulia.gubitosi@unina.it }
\and  Jos\'e Javier Relancio \at Departamento de Matemáticas y Computación \vspace{5pt}, 09001 Burgos, Spain \email{jjrelancio@ubu.es }}
\maketitle

\abstract{We review the main features of models where relativistic symmetries are deformed at the Planck scale. We cover the motivations, links to other quantum gravity approaches, describe in some detail the most studied theoretical frameworks, including Hopf algebras, relative locality, and other scenarios with deformed momentum space geometry, discuss possible phenomenological consequences, and point out current open questions.}

%\tableofcontents

\section{Introduction}
\label{sec:intro}

The proposal that local space-time symmetries might be deformed at the Planck scale was put forward in the early 2000's \cite{Amelino-Camelia:2000stu, Amelino-Camelia:2000cpa}.
The motivation was given by phenomenological studies 
on the possibility that the energy-momentum dispersion relation of  particles is deformed at the Planck scale $E_P\sim 10^{19}$GeV, such that\footnote{This formula is to be understood as indicating the lowest-order correction to the standard special-relativistic expression in powers of the particle's energy over the Planck energy, where the order is given by the positive integer $n$ and $\eta$ is a dimensionless parameter indicating the strength of the effect at the Planck scale. In general, formulas considering all-order corrections go beyond this simple power-law expression, see e.g. Section \ref{sec:kPpheno}. } \cite{Amelino-Camelia:1997ieq, Gambini:1998it, Alfaro:1999wd, Schaefer:1998zg, Biller:1998hg, Aloisio:2000cm, Amelino-Camelia:2000bxx}
\be\label{eq:MDR}
m^2 c^4=E^2-|\vec p|^2 c^2+\eta \frac{E^n}{E_P^n}|\vec p|^2 c^2\,.
\ee
Because this law is  not covariant under the standard Lorentz transformations linking inertial observers in special relativity, the first studies proposing such a modified dispersion relation assumed that invariance under Lorentz symmetries was broken at ultra-high energies, where the modification appearing in \eqref{eq:MDR} becomes relevant. As a consequence, a preferred reference frame would emerge, where the dispersion law would take the form \eqref{eq:MDR}, and this frame was  typically identified with the rest frame with respect to the cosmic microwave background. This scenario is usually called Lorentz Invariance Violation (LIV) and is discussed in another chapter of this book.

In the alternative framework proposed in \cite{Amelino-Camelia:2000stu, Amelino-Camelia:2000cpa, Amelino-Camelia:2002uql, Amelino-Camelia:2002cqb},  a modified dispersion relation of the kind of \eqref{eq:MDR} can take the same form in all inertial frames of reference, if the laws of transformation between these frames are in turn modified. In particular, they must admit two relativistic invariant quantities, the speed of light $c$ and the Planck energy $E_P$. For this reason, such framework was called Doubly Special Relativity (DSR). 

The relation between the special relativistic and the DSR scenario can be understood in analogy to the transition from the Galilean relativity of Newtonian mechanics to special relativity \cite{Amelino-Camelia:2011gae}. In Galilean relativity the (kinetic) energy of a particle is related to its momentum $\vec p\equiv m\vec v$ by
\be
E=\frac{|\vec p|^2}{2m}\,.
\ee
All observers moving with relative constant velocity see the same law, and are linked by Galilean boosts:
\be
B_j^G=i E\frac{\partial}{\partial p_j}\,.
\ee
Notice that according to these laws velocities add up linearly and there is no maximum speed.
Special relativity can be seen as a deformation of Galilean relativity that emerges when considering large velocities. In special relativity the energy-momentum dispersion relation reads
\be\label{eq:SRdisprel}
E^2=|\vec p|^2 c^2+m^2 c^4\,,
\ee
where $c$ is a velocity scale. This law is not covariant under Galilean boosts, so that, in order for it to  take the same form for all observers moving at constant relative speed, the laws of transformation linking these observers need to be deformed. These transformations, the Lorentz boosts,
\be
B_j=i \frac{p_j}{c^2}\frac{\partial}{\partial E}+i E\frac{\partial}{\partial p_j}\,,
\ee
are a deformation of the Galilean boosts that make \eqref{eq:SRdisprel} covariant. They are such that the speed scale $c$ is a relativistic invariant, identified with the speed of light, and that velocities no longer add linearly. Galilean boosts (and Galilean relativity in general) are recovered in the small velocity limit, $\frac{|\vec v|}{c}\to 0$ (see e.g. \cite{Ballesteros:2019mxi, Ballesteros:2020uxp}).

A further modification of the laws linking  observers that move at constant relative speed, which generalizes the Lorentzian boosts, allows us to retain covariance when extending the dispersion relation from the special relativistic form \eqref{eq:SRdisprel} to the modified form \eqref{eq:MDR}. Just as the extension of boosts from the Galilean form to the Lorentzian form (necessary to describe a high-velocity regime) introduces an new invariant scale $c$,  the  extension of boosts to their DSR form (supposedly necessary to describe a very-high energy regime) introduces another relativistic invariant scale, the Planck energy $E_P$.\footnote{While in special relativity $c$ is the maximum allowed speed, in DSR it is to be understood as the speed of low-energy massless particles. And the Planck energy is a relativistic invariant, but is not necessarily the maximum allowed energy. It might be the case in some specific models, but it is not true in general.} Explicit examples of these modified boost transformations are provided in Sections \ref{sec:dsr} and \ref{sec:kPpheno}.

While the basic ideas behind the DSR proposal are quite simple, the mathematical formalization  and the study of  phenomenological implications have already taken the efforts of many researchers over the past two decades. On the theoretical level, we have now several mathematical frameworks that can accommodate the DSR principles, some more developed than others. These include most notably quantum groups and Hopf algebras (see Section \ref{sec:kappa}), curved momentum space models and modified phase space models (see Section \ref{sec:relloc} and \ref{sec:comparison}). Each framework is more suited to study specific phenomenological implications, so it is worth pursuing all of them in parallel. On the phenomenological side, one of the most relevant advancements concerns the uncovering of the deep links between modified boost transformations and the loss of absolute locality. Just like the modification of boost transformations induced by the transition   from Galilean to special relativity requires us to give up the absoluteness of simultaneity, in the transition from special to doubly special relativity we are required to give up the absoluteness of locality. This was understood at the beginning of the past decade, leading to the development of the relative locality proposal, whose implications  are  the center of a very active research programme (see Section \ref{sec:rellocspacetime}).

In this review, we aim at providing the reader with the current state of the art of this research field, highlighting the progress that has been made in the theoretical modelling and the phenomenological developments, and pointing out current open questions.

\subsection{Link to more fundamental QG frameworks}\label{sec:fundamental}

As we have discussed above, DSR was motivated by phenomenological considerations relevant in searches for  effects induced by Planck-scale physics. However, subsequent studies showed that deformations of relativistic symmetries can emerge in specific limits of more fundamental quantum gravity theories.

For example, it is now well established that departures from special relativity  could arise  in a ``semiclassical'' regime of quantum gravity, where the gravitational degrees of freedom are integrated out and leave an effective field theory for the matter fields. This cannot be demonstrated explicitly for a full 3+1 dimensional quantum theory of gravity, but it was shown to be the case in 2+1 dimensions, where gravity can be quantized as a topological field theory and can be coupled to point particles, represented by topological defects   \cite{Matschull:1997du, Bais:2002ye, Meusburger:2003ta, Amelino-Camelia:2003ezw, Freidel:2003sp, Freidel:2005me, Amelino-Camelia:2012zkr, Rosati:2017toi}.  

 In the loop quantum gravity approach, one can adopt a perspective suggesting deformations of relativistic symmetries in the regime where the large-scale (coarse-grained) space-time metric is flat  \cite{Bojowald:2012ux, Mielczarek:2013rva, Amelino-Camelia:2016gfx, Cianfrani:2016ogm}. This is done by studying the  modifications to the hypersurface deformation algebra, which is the algebra describing invariance with respect to local diffeomorphisms. Such modifications provide a picture \cite{Amelino-Camelia:2016gfx} that is consistent with deformations of the relativistic symmetries of the kind that are encountered in the $\kappa$-Minkowski non-commutative spacetime, described in Section \ref{sec:kappa}.  Besides these studies, more heuristic arguments supporting the emergence of deformed relativistic symmetries  have also been put forward in the context of 3+1 dimensional loop quantum gravity \cite{Amelino-Camelia:2003ezw, Smolin:2005cz} and of polymer quantization \cite{Amelino-Camelia:2017utp}.

 Finally, it has been established that deformed relativistic symmetries emerge in the context of non-commutative space-time geometry \cite{Majid:1999tc, AmelinoCamelia:1999pm, Amelino-Camelia:1999jfz,Bruno:2001mw,  Kowalski-Glikman:2002eyl,KowalskiGlikman:2002jr, Agostini:2003vg, Amelino-Camelia:2002siu, Agostini:2006nc, Amelino-Camelia:2007yca, Amelino-Camelia:2007une, Amelino-Camelia:2007rym}. This point will be discussed in greater detail in Section \ref{sec:kappa}, for the specific case of the $\kappa$-Minkowski spacetime.

\section{Doubly special relativity - phenomenological models}
\label{sec:dsr}

The fundamental ingredients to define the phenomenology associated to a  DSR kinematical model are:
\begin{itemize}
    \item the energy-momentum dispersion relation, which can be schematically denoted as\footnote{From now on we set $c=1$.} 
\be
\mathcal C(E, \vec p)=\mu^2\,,
\ee
where $\mu$ is (a function of) the mass of the particle;
 \item 
the conservation laws of energy and momenta in interactions, which for $n$ incoming particles with momenta $p^{(i)}$ and $m$ outgoing particles with momenta $p^{(o)}$ can be written as
\beq
\left(p^{(i)}_1\oplus p^{(i)}_2\oplus ...p^{(i)}_n\right)_\mu=\left(p^{(o)}_1\oplus p^{(o)}_2\oplus ...p^{(o)}_m\right)_\mu\,,\label{eq:DSRconservation}
\eeq
were $\oplus$ encodes a deformed addition law of energy and momenta;
\item the laws of transformation between inertial observers, encoded via the action of boost transformations on energy and momenta, such that\footnote{One might also consider deformations of the other relativistic symmetries, but here we will only focus on boosts for simplicity.}
\beq
p_\mu\rightarrow (B^\xi\triangleright p)_\mu\,,
\eeq
where $\xi$ is the rapidity characterizing the magnitude of the boost.\footnote{In the following we will sometimes use a simplified notation omitting the explicit indication of the four-vector index $\mu$.}
\end{itemize}
These ingredients need to combine into quite a rigid structure, constrained by the requirement that relativistic invariance is not spoiled  \cite{Amelino-Camelia:2011gae, Carmona:2012un, Amelino-Camelia:2013sba,Carmona:2016obd}.
The observer independence of the dispersion relation can be stated as the requirement that it is invariant under boosts
\beq
 \mathcal C(B^\xi\triangleright E, B^\xi\triangleright \vec p) = \mathcal C(E, \vec p)\,.
\eeq

Covariance of the conservation laws is achieved if the sum of momenta of all incoming  (outgoing) particles transforms as a momentum under boosts:
\beq
q=p_1\oplus p_2...\oplus p_n \leftrightarrow B^\xi\triangleright q= B^\xi\triangleright \left(p_1\oplus p_2...\oplus p_n \right)\,.
\eeq
In special relativity (where momenta add linearly, $p_1\oplus p_2...\oplus p_n=p_1+ p_2...+ p_n$) the above condition is achieved by asking that
\beq
 B^\xi\triangleright \left(p_1\oplus p_2...\oplus p_n \right)=\left(B^\xi\triangleright p_1\right)\oplus \left(B^\xi\triangleright p_2\right)...\oplus\left(B^\xi\triangleright p_n \right)\,,
\eeq
and this is also the case for DSR models with a commutative law of addition of momenta $\oplus$.
However, one may have DSR models where the deformed addition rule $\oplus$ is noncommutative (a notable example is provided by models based on the $\kappa$-Poincar\'e Hopf algebra, discussed in the following Section). In this case, covariance of the conservation law \eqref{eq:DSRconservation} can only be achieved if the boost acts on  systems of interacting particles in a non-trivial way. Namely, the rapidity parameter, with which different particles participating in the interaction transform,  depends on the momenta of the other particles \cite{Gubitosi:2011hgc, Carmona:2012un,Amelino-Camelia:2013sba, Gubitosi:2019ymi}. Considering the addition of $n$ momenta $p_i$, the action of boosts is given by 
\beq
 B^\xi\triangleright \left(p_1\oplus p_2...\oplus p_n\right)\equiv \left(B^{\xi_1}\triangleright p_1\right)\oplus \left(B^{\xi_2}\triangleright p_2\right)...\oplus \left(B^{\xi_n}\triangleright p_n\right)\,,\nonumber\\ \label{eq:TotalBoost}
\eeq
where $\xi_1=\xi_1(p_2,...p_n)$, $\xi_2=\xi_2(p_1,p_3...p_n)$ and so on, such that $\xi_i=\xi$ when $p_1,...,p_{i-1},p_{i+1},...,p_n$ vanish.
This notion of covariance  of the conservation laws is a generalization of the one we are familiar with, based on the intuition we built working with special relativity. Even though this might seem counter-intuitive, it can be shown that it does not lead to the emergence of preferred observers (see a detailed discussion in \cite{Amelino-Camelia:2013sba}, arXiv version). Very recently, issues related to what is called "history problem" have emerged in association to boost actions of the sort of \eqref{eq:TotalBoost}. Since this is quite a subtle point which is currently under further study, we are not going to discuss the details here. The interested reader can refer to \cite{Gubitosi:2019ymi}.

When  considering only first order corrections to special relativity  (e.g., the $n=1$ case of \eqref{eq:MDR}), the constraints we just discussed can be translated into constraints on the coefficients of the possible correction terms that can be added to the dispersion relation, the law of addition of momenta, and the boost \cite{ Amelino-Camelia:2011gae, Carmona:2012un,Carmona:2016obd}. In particular, it can be shown that these constraints imply some ``golden rules'' on the allowed physical processes for theories without preferred frames: for example, photons cannot decay into electron-positron pairs and it must be possible for a photon of any arbitrarily low energy to produce electron-positron pairs when it interacts with a sufficiently high-energy photon \cite{ Amelino-Camelia:2011gae}. These conditions guarantee that there is no threshold associated to a photon, a fundamental requirement for any DSR theory.  In fact, because the energy of a photon could be tuned above or below the threshold with an appropriate boost, the existence of such threshold would identify a preferred frame.

Having exposed the general requirements that a DSR model needs to meet, we are going to provide one simple example, in order to see an application of the general concepts we have just exposed.
Because the subtleties emerging when the addition law $\oplus$ is non-commutative will be discussed in detail in the following Section, here we are going to consider a  1+1 dimensional DSR model with a commutative addition law. Specifically, let us consider a DSR model where all deviations from special relativity are only relevant at the first order in the ratio $\frac{E}{E_P}$ between the particles' energy and the Planck energy.
This amounts to take  $n=1$ in the dispersion relation \eqref{eq:MDR}:
\be
m^2=E^2-p^2+\eta \frac{E}{E_P}p^2\,. \label{eq:firstorderMDR}
\ee
In order for this equation to be covariant under Lorentz boosts, their action needs to be modified in such a way that energy and momentum transform non-linearly \cite{Amelino-Camelia:2010lsq,Amelino-Camelia:2011gae, Carmona:2012un}:
\beq
\begin{array}{l}\label{eq:firstorderboost}
 B^\xi\triangleright E=E+\xi p\,,\\
 B^\xi\triangleright p= p+\xi E+\xi \frac{\eta}{E_P}\left( E^2+\frac{p^2}{2}\right)\,.
\end{array}
\eeq
{\bf Exercise:} Show that the dispersion relation \eqref{eq:firstorderMDR} is not covariant under the action of standard special-relativistic boosts and verify its covariance with respect to the deformed boosts \eqref{eq:firstorderboost}.

In turn, the modified boost transformations \eqref{eq:firstorderboost} are not compatible with standard conservation laws in interactions. Considering a process with two incoming  particles $a, b$, and two outgoing particles $c, d$, a  conservation law that is covariant under the transformation \eqref{eq:firstorderboost} is
\be
\begin{array}{c}
E_a+E_b+\frac{\eta}{E_P}  p_a   p_b=E_c+E_d+\frac{\eta}{E_P}  p_c  p_d \\
 p_a+ p_b+\frac{\eta}{E_P} \left(E_a  p_b+E_b p_a\right)= p_c+ p_d+\frac{\eta}{E_P} \left(E_c p_d+E_d  p_c\right)\,.
\end{array}
\ee
This encodes a modified law of addition of energy and momenta:
\beq
\begin{array}{c}\label{eq:commutativeaddition}
E_a\oplus E_b = E_a+E_b +\frac{\eta}{E_P}  p_a   p_b\,, \\
 p_a\oplus p_b=p_a+ p_b+\frac{\eta}{E_P} \left(E_a  p_b+E_b p_a\right)\,,
\end{array}
\eeq
which is covariant assuming the following action of boosts on the interacting particles:
\beq
\begin{array}{c}
 B^\xi\triangleright \left(E_a\oplus E_b\right)=\left(B^\xi\triangleright E_a\right)\oplus \left(B^\xi\triangleright E_b\right)\,,\\
 B^\xi\triangleright \left(p_a\oplus p_b\right)=\left(B^\xi\triangleright p_a\right)\oplus \left(B^\xi\triangleright p_b\right)\,.
 \end{array}
\eeq
{\bf Exercise:} Verify that the addition law \eqref{eq:commutativeaddition} is  covariant  with respect to the deformed boosts \eqref{eq:firstorderboost}.

Notice that since we are working at the first order in $\frac{E}{E_P}$, the modified boost action and addition law are not the unique possible choices to have a relativistic picture starting from the dispersion relation \eqref{eq:firstorderMDR}. Another possibility is discussed in Section \ref{sec:kPpheno}. More thorough studies of the possibilities available at the first order can be found in  \cite{ Amelino-Camelia:2011gae, Carmona:2012un}.

In closing this section, we want to remark that, while working in momentum space suffices to study the kinematics of interactions, other possible predictions of DSR models affect the propagation of particles (e.g., one might have an induced energy dependence in the travel time of massless particles, see Section \ref{sec:pheno}). However, in order to study this kind of effects we are required to find a suitable way to describe the spacetime in which such particles propagate, or, by the least, if we want to compute the energy-dependent shift in the time of arrival of particles with different energies, we need to define a time coordinate. As we mentioned in the introduction, and will be discussed in greater detail in Section \ref{sec:rellocspacetime}, in DSR models we expect departures from the observer-independent notion of locality that applies in special relativity, and one might have that the space-time picture depends on the energy of the particle used to probe it. So defining spacetime is a highly non-trivial task, and the various mathematical frameworks that are discussed in the following can be seen as different ways to implement a space-time picture within the DSR scenario.

\section{Hopf algebras and the example of \texorpdfstring{$\kappa$}{k}-Poincar\'e}
\label{sec:kappa}

By far the most studied example of deformation of space-time symmetries is the $\kappa$-Poincaré algebra. Such a model was introduced in the early 1990s \cite{Lukierski:1991pn,Lukierski:1992dt,Majid:1994cy} and it was historically the first attempt at modifying the algebraic structure of relativistic symmetries in order to introduce a fundamental energy scale using the theory of Hopf algebras. The study of $\kappa$-deformed relativistic kinematics as a candidate for an effective description of Planck-scale physics \cite{Lukierski:1993wx,Amelino-Camelia:1997wnq,Amelino-Camelia:1999jfz} paved the way to the formulation of DSR models \cite{Amelino-Camelia:2000stu,Amelino-Camelia:2000cpa}.

\subsection{Emergence of Hopf algebra structures in quantum theory}

In order to understand how Hopf algebra structures allow for the introduction of an additional invariant scale in the description of space-time symmetries, it will be necessary to first briefly review how Hopf algebraic structures emerge in the description of symmetries in physical systems (for a more extensive and pedagogical treatment we refer the reader to \cite{Arzano:2021scz}).

In relativistic quantum theory invariance under the isometries of Minkowski spacetime requires that the states describing elementary particles carry a unitary irreducible representation of the Poincar\'e group. For a real scalar field, for example, we have a ``one-particle'' Hilbert space $\mathcal{H}$, whose elements can be given in terms of complex functions on the positive mass-shell in four-momentum space and whose elements we denote as kets labelled by the spatial momentum carried by the particle $|\vec{k}\rangle \in \mathcal{H}$. Multi-particle states will be described by (symmetrized) tensor products of such irreducible representations belonging to the Fock space \cite{Geroch:1985ci}. Let us now look at how observables act on such states. We focus on the specific example of observables $P_i$, the generators of space translations of the Poincaré algebra. One-particle states labelled by the linear momentum above are diagonal under the action of these operators
\be\label{pketk}
P_i |\vec{k}\rangle = k_i  |\vec{k}\rangle\,,
\ee
where $k_i$ is the i-th component of the vector $\vec{k}$. The action of the observable $P_i$ on generic Fock space elements is given by its {\it second quantized} version 
\begin{equation}\label{dpi}
d\Gamma(P_i)\equiv 1+ P_i + (P_i\otimes 1+1\otimes P_i)
 + (P_i\otimes 1\otimes 1 + 1\otimes P_i\otimes 1+ \ 1\otimes 1\otimes P_i)+...\,,
\end{equation}
where $1$ is the identity operator. The additional information required to extend  the representation of the Poincaré algebra from the one-particle Hilbert space to the Fock space, as encoded in \eqref{dpi}, can be formalized in terms of an operation called the {\it coproduct} 
\be\label{coprordpi}
\Delta P_i = P_i \otimes 1 + 1\otimes P_i\,,
\ee
in terms of which \eqref{dpi} can be written as 
\begin{equation}
d\Gamma(P_i)\equiv 1+ P_i + \Delta P_i + \Delta_2 P_i  + ... + \Delta_n P_i + ...\,,
\end{equation}
where
\be\label{deltan}
\Delta_n \equiv (\Delta \otimes 1)\circ \Delta_{n-1}\,,\qquad n\geq 2\,,
\ee
with $\Delta_1\equiv \Delta$. The coproduct \eqref{coprordpi} gives us the action of the observable $P_i$ linear on a two-particle state
\begin{equation}
\label{symmst}
|\vec{k}\,\vec{l}\rangle \equiv \frac{1}{\sqrt{2}}\,(|{\vec k}\rangle\otimes |{\vec l}\rangle+|{\vec k}\rangle\otimes |{\vec l}\rangle)\,.
\end{equation}
In particular
\be
\Delta P_i  |\vec{k}\,\vec{l}\rangle = (k_i+l_i) |\vec{k}\,\vec{l}\rangle\,,
\ee
and thus the eigenvalue of $\Delta P_i$ on a two-particle state is simply its total linear momentum. For our purposes it is important to notice that the coproduct encodes the property of {\it additivity of quantum numbers}. As we will see, the possibility to render {\it non-abelian} such a property for quantum numbers associated to space-time symmetries is at the core of the concept of deformation we consider.

Before proceeding we need to introduce one more ingredient concerning the action of observables on the states $\langle \vec{k}|$, i.e., on elements of the dual\footnote{By definition, elements of the dual of a Hilbert space $\mathcal{H}$ are continuous linear maps from $\mathcal{H}$ to $\mathbb{C}$. Given the inner product  $\langle \vec{k}'|\vec{k}\rangle$ on $\mathcal{H}$, it is evident that bra $\langle \vec{k}| $ is an element of the dual space.} to the one-particle Hilbert space $\mathcal{H}^*$. Since $\mathcal{H}$ carries a representation of the Poincaré algebra, on the space $\mathcal{H}^*$ one can define a {\it dual} representation. Starting from the action \eqref{pketk}, one defines the action of $P_i$ on a vector $\langle \vec{k}'| \in \mathcal{H}^*$, so that the following equality holds
\be
(P_i \langle \vec{k}'|) |\vec{k}\rangle = - \langle \vec{k}'| (P_i |\vec{k}\rangle)\,.
\ee
We thus see that the dual representation defines an action {\it from the left} of the translations generators on bras given by
\be\label{dualrp}
P_i \langle \vec{k}| = - k_i \langle \vec{k}|\,.
\ee
Notice that such action is different from the action from the right, which is simply obtained by taking the hermitian adjoint of \eqref{pketk}
\be
\langle \vec{k}| P_i \equiv (P_i |\vec{k}\rangle)^{\dagger} = \langle \vec{k}| k_i\,.
\ee
We thus see that the dual representation can be defined in terms of a map, known as the {\it antipode} 
\be
S(P_i) = -P_i\,,
\ee
connecting the left and right action of the generators $S(P_i)$ on dual states
\be\label{antipo2}
P_i \langle \vec{k}| = - k_i \langle \vec{k}| =  \langle \vec{k}| (- k_i) \equiv \langle \vec{k}| S(P_i)\,.
\ee
To understand the physical role of the antipode map let us recall that given the one-particle Hilbert space $\mathcal{H}$, the space describing anti-particles is given by the complex conjugate Hilbert space $\bar{\mathcal{H}}$.  
Such   Hilbert space is isomorphic to the dual Hilbert space $\mathcal{H}^*$ \cite{Arzano:2021scz}, and thus, for example, for a complex scalar field the bras $\langle k|$ can be identified with {\it antiparticle states}. This shows that the antipode map introduced above describes the way observables act on antiparticle states.

At the algebraic level, the coproduct and the antipode maps are additional ingredients which (together with certain consistency conditions, the interested reader can consult \cite{Arzano:2021scz} for details) equip the algebra of generators of space-time symmetries (and as a matter of fact of any quantum observable) with the structure of a Hopf algebra. We thus see that Hopf algebra structures are not just an abstract mathematical construct, but are used in our everyday quantum field theory when we look at the action of observables on antiparticles or on systems with more than one particle.

\subsection{The \texorpdfstring{$\kappa$}{k}-Poincar\'e Hopf algebra and spacetime relativistic symmetries}
After our brief detour concerning the algebraic structures underlying the action of symmetry generators on the states of a relativistic quantum system, we are now ready to introduce the $\kappa$-Poincaré Hopf algebra. The best way to understand the structure of such deformation of the Poincaré algebra is to start from a four-momentum space which is no longer a vector space, as in ordinary relativistic systems, but its geometry is that of a {\it non-abelian Lie group} which admits an action of the Lorentz group. In the $\kappa$-deformed context such group is denoted by $AN(3)$ and it is defined by the {\it Iwasawa decomposition} of the five-dimensional Lorentz group $SO(4,1)$. Such decomposition is better understood by starting from the Lie algebra $\mathfrak{so}(1,4)$ written as a direct sum of subalgebras
\begin{equation}\label{iwso41}
    \mathfrak{so} (1,4)=\mathfrak{so}(1,3)\oplus \mathfrak{n}\oplus\mathfrak{a}\,,
\end{equation}
 where the algebra $\mathfrak{a}$ is generated by the element
 \begin{equation}\label{3.19}
    H  =\left[\begin{array}{ccccc}
    0&0 &0 &0 & 1 \\
    0 & 0 & 0 & 0 & 0 \\
    0 & 0 & 0 & 0 & 0 \\
    0 & 0 & 0 & 0 & 0 \\
    1& 0 & 0 & 0 & 0 \
  \end{array}\right]\,,
\end{equation}
and the algebra $\mathfrak{n}$ by the elements
\begin{equation}\label{3.20}
  \mathfrak{n}_i
    =\left[ \begin{array}{ccc}
 0 & (\epsilon_i)^T & 0 \\
 \epsilon_i  & 0  & \epsilon_i \\
    0 & -(\epsilon_i)^T & 0
\end{array}\right]\,,
\end{equation}
where $\epsilon_i$ are unit vectors in i-th direction ($\epsilon_1 = (1,0,0)$, etc), and T denotes transposition. 

We now introduce a constant parameter, which carries dimensions of energy, denoted by $\kappa$. We can use such constant to define non-commuting objects with dimension of length\footnote{As we will explain in the following section, in some approaches to DSR based purely on the geometry of momentum space one assumes to be in a ``semiclassical'' regime of quantum gravity, such that the Planck constant $\hbar$ and the Newton constant $G$ vanish, but their ratio is fixed and finite. In this regime, one can build an energy scale $E_P$ but not a length scale $L_P\to 0$. In the context of Hopf algebra and non-commutative geometry, this is not  the regime that is considered, since one needs a constant with dimensions of length to govern space-time noncommutativity as in \eqref{kminkn}.}
\be
X^0 = -\frac{i}{\kappa}\, H\,,\qquad X^i = \frac{i}{\kappa}\, \mathfrak{n}_i\,.
\ee
These {\it non-commuting coordinates} obey the commutator
	\beq \label{kminkn}
	[X^0, X^i] = \frac{1}{\kappa} X^i\,, \qquad [X^i, X^j] = 0\,,
	\eeq
known in the literature as the $\kappa$-Minkowski non-commutative spacetime \cite{Majid1994}.

From \eqref{iwso41} it follows that every element $\lambda$ of the group $SO(1,4)$ can be decomposed as follows
\be
 \lambda=(K n a)\,,\quad \mbox{or} \quad \lambda=(K\,\vartheta\,n a)\,,
\ee
where $K\in SO(1,3)$, the element $a$ belongs to group $A$ generated by $H$
\begin{equation}\label{3.22}
    A=\exp\left( i {k_0}X^0 \right) =\exp\left( \frac{k_0}\kappa H \right)=\left[\begin{array}{ccccc}
    \cosh \frac{k_0}{\kappa}&0 &0 &0 & \sinh \frac{k_0}{\kappa}\\
    0 & 1 & 0 & 0 & 0 \\
    0 & 0 & 1 & 0 & 0 \\
    0 & 0 & 0 & 1 & 0 \\
    \sinh \frac{k_0}{\kappa} & 0 & 0 & 0 & \cosh \frac{k_0}{\kappa} \
  \end{array}\right]\,,
\end{equation}
and the element $n$ belongs to group $N$ generated by the matrices $\mathfrak{n}_i$
\begin{equation}\label{3.35}
    N=\exp\left(ik_i X^i\right)=\exp\left(-\frac{1}{\kappa} k_i \mathfrak{n}_i\right)=\left[ \begin{array}{ccccc}
  1+\frac{1}{2\kappa^{2}}\vec{k}^{\;2}& \frac{k_{1}}{\kappa} & \frac{k_{2}}{\kappa}  & \frac{k_{3}}{\kappa} &  \frac{1}{2\kappa^{2}}\vec{k}^{\;2}\\
    \frac{k_{1}}{\kappa} & 0 & 0 & 0 & \frac{k_{1}}{\kappa} \\
    \frac{k_{2}}{\kappa} & 0 & 0 & 0 & \frac{k_{2}}{\kappa} \\
    \frac{k_{3}}{\kappa} & 0 & 0 & 0 & \frac{k_{3}}{\kappa} \\
    -\frac{1}{2\kappa^{2}}\vec{k}^{\;2}& -\frac{k_{1}}{\kappa} & -\frac{k_{2}}{\kappa} & -\frac{k_{3}}{\kappa} & 1-\frac{1}{2\kappa^{2}}\vec{k}^{\;2} \
  \end{array}\right]\,,
\end{equation}
and $\vartheta$=diag(-1,1,1,1,-1). 

The group $AN(3)$ can be identified with the product group $N A$. Elements of $AN(3)$ can be expressed as {\it plane waves} on the non-commutative $\kappa$-Minkowski spacetime with the time component appearing to the right \cite{Amelino-Camelia:1999jfz}
\be\label{plankap}
   \hat{e}_k = \exp(ik_{i}X_{i})\exp(ik_{0}X^{0})\,.
\ee

The four-momenta $k_{\mu}$ are coordinate functions on the group $AN(3)$ known as {\it bicrossporduct coordinates}. The quotient group structure of $AN(3)\simeq SO(1,4)/SO(1,3)$ allows us to obtain another system of coordinates. Indeed, it is well known that the quotient of Lie groups $SO(1,4)/SO(1,3)$ is the de Sitter space. Acting with the subgroup $AN(3)$ of $SO(1,4)$ on the vector $\mathcal{O}$
\begin{equation}\label{3.24}
 {\cal O}   =\left[\begin{array}{c}
    0 \\
    0 \\
    0 \\
    0 \\
    \kappa \
  \end{array}\right]\,,
\end{equation}
 we obtain the coordinates 
$$
 p_{0}=\kappa\sinh\frac{k_{0}}{\kappa} +\frac{\vec{k}{}^{2}}{2\kappa}
 e^{\frac{k_{0}}{\kappa}}\,,
$$
$$
 p_{i}=k_{i}e^{\frac{k_{0}}{\kappa}}\,,
$$
\begin{equation}\label{3.25}
 p_{4}=\kappa\cosh\frac{k_{0}}{\kappa} -\frac{\vec{k}{}^{2}}{2\kappa}
 e^{\frac{k_{0}}{\kappa}}\,,
\end{equation}
on de Sitter space defined as the submanifold of the five-dimensional Minkowski space by the equation
$$
-p_{0}^2 + p_{i}^2 + p_{4}^2 = \kappa^2\,.
$$
{\bf Exercise:} Using the relations \eqref{3.25} show that the coordinates $k_0,\vec k$ correspond to comoving coordinates on the de Sitter manifold, according to which the line element is $ds^2= dk_0^2-e^{2 k_0/\kappa}d\vec k^2$. Use \cite{Gubitosi:2011hgc} as guidance.

Notice how such {\it embedding} coordinates only cover “half'' of de Sitter manifold determined by the inequality
\be
p_0+p_4>0\,.
\ee
Considering two sets of coordinates
$$
 p_{0}=\pm (\kappa\sinh\frac{k_{0}}{\kappa} +\frac{\vec{k}{}^{2}}{2\kappa}
 e^{\frac{k_{0}}{\kappa}})\,,
$$
$$
p_{i}=\pm k_{i}e^{\frac{k_{0}}{\kappa}}\,,
$$
\begin{equation}\label{3.29}
 p_{4}=\mp(\kappa\cosh\frac{k_{0}}{\kappa} -\frac{\vec{k}{}^{2}}{2\kappa}
 e^{\frac{k_{0}}{\kappa}})\,,
\end{equation}
we can cover the entire de Sitter manifold, and this is reflected in the fact that the group SO(1,4) can be written in the form
\begin{equation}\label{3.30}
    SO(1,4)=KNA \cup K\vartheta NA\,.
\end{equation}

In fact, non-trivial geometrical properties of momentum space are a generic feature of DSR models. The relative locality proposal, see Section \ref{sec:relloc}, takes this observation as a starting point.

We are now ready to discuss the algebra structure of $\kappa$-deformed symmetries. We start with translation generators that can be defined as acting like ordinary derivatives on non-commutative plane waves, once we have chosen an ordering of the non-commuting factors \cite{Agostini:2003vg}. Let us focus on the time-to-the-right ordered plane waves \eqref{plankap} and define translation generators associated to bicrossproduct coordinates as
\be\label{bcpcoord}
\tilde{P}_{\mu}\, \hat{e}_k \equiv k_{\mu} \hat{e}_k\,.
\ee
The eigenvalues $k_{\mu}$ are coordinates on the $AN(3)$ group manifold, and thus, just real numbers, from which we can deduce that the generators of translations commute
\be
[\tilde{P}_{\mu}, \tilde{P}_{\nu}] = 0\,.
\ee
In order to determine the other commutators of the $\kappa$-Poincaré algebra we need to define the action of the Lorentz group on the $AN(3)$ momentum space. 

Let us introduce the following notation for the Iwasawa decomposition for an element of $\lambda \in SO(4,1)$
\be
\lambda = K_g\, g\,,
\ee
with $K_g\in SO(3,1)$ and $g\in AN(3)$. Uniqueness of the Iwasawa decomposition guarantees that given the Lorentz group element $K_g$ and $g$, there are unique elements $K'_{g'}$, $g'$, satisfying 
\begin{equation}\label{a2}
   K_g\, g = g'\, K'_{g'} \,,
\end{equation}
so that one {\it defines} the Lorentz transformed group valued momentum as
\begin{equation}\label{a3}
   g'\,= K_g\, g  (K'_{g'})^{-1}\,. 
\end{equation}
In order to derive the action of such Lorentz transformation on momenta, and thus the commutators with the generators of translations, one can write the following expression for infinitesimal transformations
\begin{equation}\label{kginf}
    K_g \approx 1+i \xi^a \mathfrak{k}_a\,,\quad K'_{g'} \approx 1+i \xi^a \, h_a^b(g) \mathfrak{k}_b\,,
\end{equation}
where $\mathfrak{k}_a$ are the generators of the Lorentz algebra $\mathfrak{so}(1,3)$, and $h_a^b(g) $ is a matrix function of the momentum.
We can write the momentum group element as a matrix in terms of embedding coordinates \eqref{3.24}
 \begin{equation}\label{gembed}
     g=\left(
  \begin{array}{ccc}
\tilde p_4& \mathbf{p}/p_+& p_0\\
\mathbf{p} & 1 & \mathbf{p}\\
\tilde p_0&-\mathbf{p}/p_+&p_4\\
  \end{array}
\right)\,,
 \end{equation}
where $1$ is the unit $3\times3$ matrix and $p_+ = p_0 + p_4$. Plugging such a matrix expression for $g$ and $g'$, and \eqref{kginf} in \eqref{a3}, one remarkably obtains \cite{Arzano:2022ewc} that the action of the Lorentz generators (and thus also of the Lorentz group) on the four-momenta $p_{\mu}$ is the {\it ordinary one}. Thus, defining the set of translation generators associated to embedding coordinates through the following action on plane waves 
\be
P_{\mu}\, \hat{e}_k \equiv p_{\mu} \hat{e}_k\,,
\ee
(notice how this differs from \eqref{bcpcoord}) we have that they obey the ordinary commutators with generators of rotations $M_i$ and $N_i$:
\begin{align}\label{Lorentz transalgem}
    [M_i, P_j] &= i\, \epsilon_{ijk} p_k, \quad [M_i, P_{0}] =0\nonumber\\
   \left[N_{i}, {P}_{j}\right] &= i\,  \delta_{ij} P_0 ,\,\,\, \left[N_{i},P_{0}\right] = i\
P_{i}\,.
\end{align}
Now we come to one of the key points which distinguishes the $\kappa$-deformed algebra from the standard Poincaré algebra. As the reader might have noticed, we have already introduced two different sets of translation generators: the ones associated to bicrossproduct coordinates, which we denoted by $\tilde{P}_{\mu}$, and those associated to embedding coordinates $P_{\mu}$. In the literature, these different sets of generators are known as different {\it bases} of the $\kappa$-Poincaré algebra. The generators $P_{\mu}$ (together with the generators of rotations and boosts) are known as the “classical basis'' of the $\kappa$-Poincaré algebra \cite{Borowiec2010}, since at the Lie algebra level they just reproduce the standard Poincaré algebra. The generators $\tilde{P}_{\mu}$ determine the so-called “bicrossproduct basis'' \cite{Majid1994}. Using the coordinate transformation \eqref{3.25}, one can easily see that the commutators between the bicrossproduct generators of translations and the Lorentz ones are {\it deformed} 
\begin{align}\label{Lorentz transalg}
    [M_i, \tilde{P}_j] &= i\, \epsilon_{ijk} \tilde{P}_k, \quad [M_i, \tilde{P}_{0}] =0\,,\nonumber\\
   \left[N_{i}, \tilde{P}_{j}\right] &= i\,  \delta_{ij}
 \left( {\frac\kappa2 } \left(
 1 -e^{-2\tilde{P}_{0}/\kappa}
\right) + {\frac{\mathbf{\tilde{P}}^2}{2\kappa}}  \right) - i\,
\frac1\kappa\, \tilde{P}_{i} \tilde{P}_{j} ,\,\,\, \left[N_{i},\tilde{P}_{0}\right] = i\
\tilde{P}_{i}\,.
\end{align}
{\bf Exercise:} compute the commutators \eqref{Lorentz transalg} starting from \eqref{Lorentz transalgem} and using the change of basis \eqref{3.25}.

Notice how in the classical basis since the algebra is undeformed so is the mass Casimir invariant
\be
\mathcal{C}_{\kappa}(P)=
P^2_0-\vec{P}^2 \,,
\ee
reflecting the invariance under $SO(3,1)$ transformations of the subspaces $p_4=const.$ of de Sitter space. Of course, the same invariant object written in bicrossproduct coordinates will have a very complicated non-linear form. Thus, we see that the curved manifold structure of momentum space makes it possible to have non-linear energy-momentum dispersion relations, which at leading order are formally analogous to the modifications of the energy-momentum dispersion relations characterizing models in which Lorentz invariance is broken. Here, however, such modifications are fully compatible with the action of Lorentz transformations which are now deformed according, e.g., to the modified commutators \eqref{Lorentz transalg}. This feature is at the basis of the general ideas of DSR models, in which the deformation parameter $\kappa$ is seen as a fundamental, observer independent, Planckian energy scale, see Sections \ref{sec:intro}, \ref{sec:dsr} and \ref{sec:kPpheno}. The price to pay for the introduction of such a scale in a way in which Poincaré symmetries are preserved is to renounce to the Abelian additivity of quantum numbers associated to such symmetries, as we show below. It is important to notice that this last feature is also present in the classical basis. So this basis is not equivalent to special relativity, despite having a trivial algebra and Casimir.

Before proceeding, let us summarize the results obtained so far: in the classical basis, the $\kappa$-Poincaré algebra is nothing but the ordinary Poincaré algebra. In the bicrossproduct basis, the commutators are given by \cite{Arzano:2021scz}
\begin{align}\label{kpoinc}
[\tilde{P}_{\mu}, \tilde{P}_{\nu}] & = 0\,,\\
    [M_i, \tilde{P}_j] &= i\, \epsilon_{ijk} \tilde{P}_k, \quad [M_i, \tilde{P}_{0}] =0\,,\nonumber\\
   \left[N_{i}, \tilde{P}_{j}\right] &= i\,  \delta_{ij}
 \left( {\frac\kappa2 } \left(
 1 -e^{-2\tilde{P}_{0}/\kappa}
\right) + {\frac{\mathbf{\tilde{P}}^2}{2\kappa}}  \right) - i\,
\frac1\kappa\, \tilde{P}_{i} \tilde{P}_{j} ,\,\,\, \left[N_{i},\tilde{P}_{0}\right] = i\
\tilde{P}_{i}\,,\\
[M_i, M_j] &= i\, \epsilon_{ijk} M_k ,\,\,\,    [M_i, N_j] = i\,\epsilon_{ijk} N_k ,\,\,\, [N_i, N_j] = -i\, \epsilon_{ijk} M_k\,,
\end{align}
while the Casimir invariant is given by 
 \be
 \mathcal{C}_{\kappa}(\tilde{P}) = 
\left(2\kappa\sinh\frac{\tilde{P}_{0}}{2\kappa}\right)^2 -\vec{\tilde{P}}{}^{2} e^{\frac{\tilde{P}_{0}}{\kappa}}\,. \label{eq:kPCasimir}
 \ee
Notice that the relation between the classical basis Casimir, $\mathcal{C}_{\kappa}(P)$, and the bicrossproduct one, $\mathcal{C}_{\kappa}(\tilde{P})$, is \cite{Arzano:2014jfa} $\mathcal{C}_{\kappa}(P)= \mathcal{C}_{\kappa}(\tilde{P})\left(1+\frac{1}{4 \kappa^2} \mathcal{C}_{\kappa}(\tilde{P}) \right)$.
As a matter of fact, the presence of the invariant energy scale $\kappa$ in the model renders any function of $\mathcal{C}_{\kappa}(P)$ a good candidate for the invariant mass Casimir. Historically, (a variant of) the bicrossproduct basis Casimir $\mathcal{C}_{\kappa}(\tilde{P})$ was first derived in the literature using contraction techniques on a q-deformed anti-de Sitter algebra \cite{Lukierski:1992dt}. For this reason, the vast majority of works focusing on the applications of the $\kappa$-Poincaré algebra adopted such Casimir to define a $\kappa$-deformed energy-momentum dispersion relation.

We are now ready to discuss the so-called {\it co-algebra} sector of the $\kappa$-Poincaré algebra, namely, the generalization to the $\kappa$-deformed setting of the coproduct and antipode maps discussed at the beginning of this section. As we have seen, working in the classical basis one can establish a Lie algebra isomorphism between the $\kappa$-Poincaré algebra and the ordinary Poincaré algebra. Thus, at the one-particle level, irreducible representations of the $\kappa$-Poincaré algebra can be identified with those of the ordinary Poincaré algebra \cite{Ruegg:1994bk}. As in ordinary quantum field theory, such irreducible representations for a scalar field can be constructed starting from plane waves. In the $\kappa$-deformed case, we deal with non-commutative plane waves \eqref{plankap}
which, as in the standard case, can be put in correspondence with kets labelled by the eigenvalues associated to space-time translation generators. States characterized by on-shell momenta provide irreducible representations of the Lie algebra. We focus on bicrossproduct generators, 
\be
\tilde{P}_{\mu} |k\rangle = k_{\mu} |k\rangle\,.
\ee
Such kets can be put in correspondence with {\it ordinary} plane waves
\be
\langle x| k\rangle \sim e_k\,
\ee
equipped with a non-commutative $\star$-product, such that 
\be
e_k \star e_l \equiv \hat{e}_k \hat{e}_l\,, 
\ee
where $\hat{e}_k$ and $\hat{e}_l$ are the ordered plane waves \eqref{plankap}. To any choice of ordering it will correspond a different choice of $\star$-product for ordinary plane waves \cite{Agostini:2003vg}. The non-commutative nature of the $\star$-product is simply a reflection of the non-Abelian structure of momentum space. Indeed, the product of non-commutative plane waves $\hat{e}_k$ and $\hat{e}_l$, results in an ordered plane wave
\be
\hat{e}_k \hat{e}_l = \hat{e}_{k \oplus l}\,, \label{eq:groupmultiplication}
\ee
where $k \oplus l$ is a non-Abelian addition law for the four-momenta $k$ and $l$. Its form can be derived explicitly (see e.g. \cite{Kosinski:1999dw}) by re-ordering the factors in the product $\hat{e}_k \hat{e}_l$ in such a way to restore the normal ordering with all the factors containing $X^0$ to the right using the Baker-Campbell-Hausdorff formula for the Lie algebra \eqref{kminkn}. The resulting addition law is 
\be
k \oplus l = (k_0 + l_0, \vec{k} + e^{-k_0/\kappa}\, \vec{l})\,.\label{eq:addlawfromgroup}
\ee
Another way to derive such a non-Abelian addition law would be to read-off the bicrossproduct four-momentum of the product matrix $\hat{e}_{k \oplus l} = \hat{e}_k \hat{e}_l$ obtained using the matrix expressions \eqref{3.22} and \eqref{3.35}. Looking at the plane wave $\hat{e}_{k \oplus l}$, we can write 
\be
\hat{e}_{k \oplus l} \sim \langle x| k \oplus l \rangle = \star (\langle x| k\rangle \otimes \langle x| l\rangle)\,,
\ee
where the $\star$-product is seen as a map defined on the tensor product of two copies of the space of functions on Minkowski spacetime. Now, since $\tilde{P}_{\mu}\hat{e}_{k \oplus l} = (k \oplus l) \hat{e}_{k \oplus l}$, we can derive the co-product for the bicrossproduct translation generators through the identity \cite{Guedes:2013vi}
\be
\tilde{P}_{\mu} \langle x| k \oplus l \rangle = \star (\langle x| (\Delta \tilde{P}_{\mu} (|k\rangle \otimes l\rangle))\,,
\ee
obtaining
\be\label{coprodbicr}
\Delta \tilde{P}_0 = \tilde{P}_0 \otimes 1 + 1\otimes \tilde{P}_0\,, \qquad \Delta \tilde{P}_i = \tilde{P}_i \otimes 1 + e^{-\tilde{P}_0/\kappa}\otimes \tilde{P}_i\,.
\ee
This shows that, in the $\kappa$-deformed context, the non-Abelian composition law of four-momenta is translated into a non-tivial coproduct for translation generators. This in turn can be seen as a non-Abelian generalization of the Leibniz rule for the action of such generators on tensor product states. The non-trivial structure of the $\kappa$-deformed coproduct is intimately related to the non-Abelian product of two momentum group elements. The operation of taking the {\it inverse} of a momentum group element translates instead into a deformation of the antipode map. In terms of non-commutative plane waves, we can define a new plane wave labelled by a new momentum $\ominus k$ such that 
\be
\hat{e}_{k} \hat{e}_{\ominus k} = 1\,,
\ee
in other words, $\hat{e}_{\ominus k} \equiv (\hat{e}_{k})^{-1}$. It is easy to see that from the definitions above we have $k \oplus (\ominus k) = 0$. One can easily show (e.g. by inverting the matrix expression for $\hat{e}_{k}$ and reading off the coordinates) that
\be
\ominus k = (-k_0, - e^{k_0/\kappa}\, \vec{k})\,.
\ee
Recalling our definition of antipode map \eqref{antipo2}, this immediately reflects on the non-trivial antipode map on translation generators
\be
S(\tilde{P}_{\mu})  = (-\tilde{P}_0, - e^{\tilde{P}_0/\kappa}\, \vec{\tilde{P}})\,.
\ee
The last ingredients needed to complete our ‘‘derivation'' of the $\kappa$-Poincaré Hopf algebra are the co-products and antipodes for the Lorentz generators. In analogy with the generators of translations, the coproducts for the Lorentz generators can be obtained from the action of the $SO(3,1)$ group on the product of two momentum $AN(3)$ group elements (an alternative derivation in terms of the so-called Weyl maps can be found in \cite{Agostini:2003vg}). Such action is given by a generalization of \eqref{a3}
\begin{equation}\label{s20}
    K_g\, g\,h\, K'^{-1}_{(gh)'} = (gh)'.
\end{equation}
We immediately see  that $(gh)'\neq g'\, h'$, which shows that the Lorentz group action on momentum space is not Leibnizian. For the antipodes, one looks at the action of Lorentz transformations on inverse group elements, namely
\be
(g^{-1})' = K'_{g'}\, g^{-1}\, K^{-1}_g\,.
\ee
Using the infinitesimal form of the transformations \eqref{kginf}, together with matrix representation of the $AN(3)$ group elements and of the Lorentz generators, one can show (see \cite{Arzano:2022ewc} for details) that the coproduct and antipode for rotation generators remain trivial
\be
\Delta M_i = M_i \otimes 1 + 1 \otimes M_i\,, \qquad S(M_i) = -M_i\,,
\ee
while for boost generators one finds
\be
\Delta(N_{i}) =  N_{i}\otimes 1+P_{+}^{-1}\otimes N_{i}+\epsilon_{ijk}\,\frac{1}{\kappa}P_{j}P_{+}^{-1}\otimes M_{k}\,,
\ee
\be
S(N_{i}) =  -N_{i}P_{+}+\epsilon_{ijk}\,\frac{1}{\kappa}P_{j}M_{k}\,,
\ee
where $P_{+} = P_0 + P_4$, with $P_{4} = \sqrt{\kappa^2 + P_{0}^2 - \vec{P}^2}$. Written in terms of the bicrossproduct generators such coproduct and antipode read
\be
\Delta (N_i)=N_i \otimes 1 +e^{-\tilde{P}_0/\kappa}\otimes N_i +\frac1\kappa\, \epsilon_{ijk} \tilde{P}_j \otimes M_k\,, \label{eq:kpboostcoproduct}
\ee
\be
S(N_i)=-e^{\frac{\tilde{P}_0}{\kappa}}(N_i-\frac{1}{\kappa}\epsilon_{ijk}\tilde{P}_jM_k)\,.
\ee
These complete our description of the co-algebra structure of the $\kappa$-Poincaré Hopf algebra.

{\bf Exercise:} Using the change of basis \eqref{3.25}, compute the co-algebra structure in terms of the generators $P_\mu$ of the classical basis.

\subsection{Link to DSR phenomenological models} \label{sec:kPpheno}

The Hopf algebra structure just described can be used to define a phenomenological model for DSR of the kind discussed in Section \ref{sec:dsr}.
In fact, because in the bicrossproduct basis the translation generators form a Hopf-subalgebra, they can be represented as an algebra of functions over momentum space \cite{Gubitosi:2011hgc, Amelino-Camelia:2011uwb, Kowalski-Glikman:2003qjp}, such that the  translation generators correspond to the coordinate functions $p_\mu$,
\beq
\tilde P_\mu(p)= p_\mu\,.
\eeq

%\footnote{Physical momenta are associated to the conserved charges under spacetime translation transformations. The issue was investigated in \cite{Agostini:2006nc, Amelino-Camelia:2007rym, Amelino-Camelia:2007une, Amelino-Camelia:2007yca}.}

Then, the Casimir \eqref{eq:kPCasimir} can be used to read a deformed dispersion relation. Given that the Casimir is by definition an invariant of the symmetry generators, it can be equated to (a function of) the squared mass of the particle $\mu^2$, so that the energy-momentum dispersion relation reads
\beq
\mu^2= \left(2\kappa\sinh\frac{{p}_{0}}{2\kappa}\right)^2 -\vec{{p}}{}^{2} e^{\frac{{p}_{0}}{\kappa}}\simeq p_0^2-\vec p^2- \frac{1}{\kappa}p_0\vec p^2 \,,
\eeq
where we have also indicated the first-order expansion in $\frac 1 \kappa$. In this scenario, the parameter $\kappa$ gives the relativistically invariant energy scale.
Invariance of this dispersion relation can be verified  explicitly by performing a boost transformation in the $j$ direction according to:\footnote{We adopt a semiclassical approximation, so that symmetry generators act on the momentum space coordinates via Poisson brackets. The properties of the generators of the Hopf algebra are inherited by the Poisson brackets  with the convention that, if  $[G, f(P_{\mu})]= i h(P_{\mu})$, then $\{G,f(p_{\mu})\} = h(p_{\mu})$, for any generator $G$ of the Hopf algebra. The functions $f$, $h$, take as argument  the translation generators $P_{\mu}$ in the first case, and the momentum space coordinates $p_{\mu}$ in the second one. This approximation is justified in the ``semiclassical'' limit we mentioned in the previous footnote and further described in Section \ref{sec:relloc}.} 
\beq
\begin{array}{l}
B^{\xi}_j \triangleright p_{0} \equiv p_0 + \xi\{N_j,p_0\} = p_0 + \xi p_j\,,\nonumber \\
B^{\xi}_j \triangleright p_{i} \equiv p_i+ \xi\{N_j,p_i\} = p_i + \xi \delta_{ij} \left[\frac{\kappa}{2}\left(1-e^{-2p_0/\kappa}\right) - \frac{1}{2\kappa}\vec p^2\right]-\xi \frac{1}{\kappa} p_i p_j\,, \label{eq:boosted_momenta}
\end{array}
\eeq
in which $\xi$ is the rapidity parameter.

When comparing the first-order expansion with \eqref{eq:firstorderMDR}, we see that the two expressions are equivalent upon setting $\frac 1\kappa=\frac{\eta}{ E_P}$. However, we are now going to show that the other ingredients of the DSR model inspired from the $\kappa$-Poincar\'e Hopf algebra, namely, the composition law and the deformed boosts, when considered at the first order in the deformation parameter, are not the same as the ones of the model discussed in Section \ref{sec:dsr}, despite the two models sharing the same first-order dispersion relation. This is possible because when working to the first order in the deformation parameter the relativistic constraints leave open some degrees of freedom in the definition of the model \cite{Amelino-Camelia:2011gae, Carmona:2012un}, so that starting from the same deformed dispersion relation one can construct different relativistic models. The model we are discussing in this section is in principle valid to all orders in $\frac{1}{\kappa}$, even though it might still be the case that it only describes physics in a limited energy range (one could imagine that Nature is such that DSR models only describe the relativistic symmetries in a limited energy range, above which one might have a full breakdown of symmetries, or a restoration of special relativistic symmetries).

Other structures of the Hopf sub-algebra of translations are linked to  the properties of the momentum space. Specifically, a deformed composition law of momenta is read off from the Hopf algebra coproduct:
\beq
\Delta(P_\mu)(p,q)=\left(p\oplus q\right)_\mu\,,
\eeq
so that the deformed addition law $\oplus$ reads
\beq
\begin{array}{c}
E_a\oplus E_b = E_a+E_b\,, \\
\vec p_a\oplus \vec p_b=\vec p_a+ e^{E_a/\kappa}\vec p_b\,.\label{eq:kPconslaw}
\end{array}
\eeq
Notice that this is also compatible with the interpretation of the momentum space as a group manifold, so that the momentum composition  is defined by the group multiplication law \eqref{eq:groupmultiplication}-\eqref{eq:addlawfromgroup}.

In contrast to the example considered in Section \ref{sec:dsr}, here we have a  composition law for spatial momenta that is noncommutative (because of the noncocommutativity of the coproduct) and associative (because of the coassociativity of the coproduct). 

As we already mentioned briefly in Section \ref{sec:dsr}, when the addition rule $\oplus$ is noncommutative then the  momenta of each particle are boosted with  rapidities that depend on the other particles' momenta, and this guarantees covariance of the addition law.
The model we are describing in this section is an example of such behaviour. \footnote{We are here going to discuss only the 1+1 dimensional case, since in 3+1 dimensions a nontrivial interplay between boost and rotation transformations occurs, such that pure boost transformations are not allowed. For details see \cite{Gubitosi:2011hgc, Amelino-Camelia:2007yca}.}
In fact, it is now well understood \cite{Gubitosi:2011hgc, AmelinoCamelia:2011yi, Gubitosi:2019ymi} that the addition law \eqref{eq:kPconslaw} is not covariant if each momentum is boosted with the same rapidity $\xi$:
\be
p_a\to B^{\xi} \triangleright p_a\,,\quad p_b\to B^{\xi} \triangleright p_b \,.
\ee
In fact, calling $q\equiv p_a\oplus p_b$, it can be shown that [{\bf Exercise}]:
\be
B^{\xi} \triangleright q \neq (B^{\xi}  \triangleright p_a) \oplus (B^{\xi}  \triangleright p_b)\,.
\ee

What does work to achieve covariance of the addition law is to account for a ``backreaction'' of the individual momenta onto the  rapidity $\xi$ \cite{Gubitosi:2011hgc, Gubitosi:2019ymi}. This is such that the rapidity with which the  second momentum transforms is affected by the first momentum:\footnote{A more general expression applies when considering finite transformations \cite{Gubitosi:2011hgc}; however, here we  only discuss the first order in $\xi$.}
\be
B^{\xi} \triangleright q =  (B^{\xi}  \triangleright p_a) \oplus (B^{\xi\triangleleft p_a}  \triangleright p_b)\,,\label{eq:backreaction}
\ee
where $\xi\triangleleft p_a \equiv e^{-E_a/\kappa}\xi$. As discussed in detail in \cite{Amelino-Camelia:2013sba}, such backreaction does not identify a preferred frame of reference and is fully compatible with relativistic invariance. {\bf Exercise:} Verify the covariance of the conservation law \eqref{eq:kPconslaw} under the transformation \eqref{eq:backreaction}. Use \cite{Gubitosi:2019ymi} as guidance.

Since the model we are discussing in this section is linked to a Hopf algebra, we can understand such an action of boosts on composed momenta in terms of the properties of the coproduct of the boost generator \eqref{eq:kpboostcoproduct}. In fact, one can interpret the backreaction \eqref{eq:backreaction} in terms of a law of ``addition'' of boost generators that dictates how composed momenta transform. The  ``total boost'' generator 
\begin{align}
N_{[p_a\oplus p_b]} = N_{[p_a]} + e^{-E_a/\kappa}N_{[p_b]},\label{eq:total_boost}
\end{align}
is defined by the coproduct of the boost generator in the underlying Hopf algebra, Eq. \eqref{eq:kpboostcoproduct}. Here, the notation $N_{[p_a]}$ indicates that the transformation only acts on $p_a$ and not on $p_b$. The ``total boost'' of rapidity $\xi$ then has the following action on the momenta of each of the two interacting particles:
\beq
\begin{array}{l}
B^{\xi} \triangleright p_a = p_a+\xi \{N_{[p_a\oplus p_b]},p_a\}= p_a+\xi \{N_{[p_a]},p_a\}\,,\nonumber \\
B^{\xi} \triangleright p_b= p_b+\xi \{N_{[p_a\oplus p_b]},p_b\}= p_b+\xi e^{-E_a/\kappa} \{N_{[p_b]},p_b\}= p_b+(\xi\triangleleft p_a) \{N_{[p_b]},p_b\}\,.
\end{array}
\eeq

When considering the composition of several momenta, the rapidity acting on each of them receives a backreaction from all the other  momenta that come before it in the composition law. Specifically, considering the addition of $n$ momenta
\be
p^{(1)}\oplus...\oplus p^{(n)}\,,
\ee
the rapidity with which the particle with momentum $p^{k}$ is boosted reads:
\be
\xi^{[p^{(k)}]}=\xi\triangleleft p^{(1)}\triangleleft ... \triangleleft p^{(k-1)}\equiv \xi\triangleleft (p^{(1)}\oplus ... \oplus p^{(k-1)})\,.
\ee
Again, we can interpret the backreaction in terms of the action of a total boost. Boosting each momentum with its own boost generator, $N_{[p^{(i)}]}$, and incorporating the backreaction on the rapidity, as explained above, is completely equivalent to boosting each momentum  with the total boost generator \cite{Gubitosi:2019ymi}
\begin{align}\label{eq:totalboost}
N_{[\bigoplus_{i=1}^n p^{(i)}]} = N_{[p^{(1)}]} + e^{-E^{(1)}/\kappa}N_{[p^{(2)}]} +  \dots + e^{-\left(\sum_{i=1}^{n-1}E^{(i)}_0\right)/\kappa}N_{[p^{(n)}]} \,.
\end{align}

\subsection{DSR on curved spacetime - the \texorpdfstring{$\kappa$}{k}-(Anti) de Sitter example}

Most of the currently available DSR models, including the $\kappa$-Poincar\'e algebra described in this section, describe deformations of the relativistic transformations in flat spacetime. However, as we will discuss in Section \ref{sec:pheno}, the most interesting phenomenological applications refer to the propagation of particles over cosmological distances, where the flat spacetime approximation is no longer valid. 

Phenomenological studies aimed at extending DSR models to de Sitter or even Friedmann-Robertson-Walker spacetimes have been undertaken in relatively recent times \cite{ Rosati:2015pga, Amelino-Camelia:2012vzf,Relancio:2020zok,Relancio:2020rys}, while some first exploratory studies were already performed more than a decade ago \cite{Amelino-Camelia:2003ezw, Marciano:2010gq}. An interesting line of investigation concerns the generalization of the  $\kappa$-Poincar\'e Hopf algebra to allow for a non-vanishing cosmological constant $\Lambda$. This  leads to a quantum-deformed (Anti)-de Sitter Hopf algebra, known as $\kappa$-(A)dS \cite{Ballesteros_1994, Ballesteros:2016bml} and its associated non-commutative spacetime \cite{Ballesteros:2019hbw,Pfeifer:2021tas}.

These investigations agree on the fact that, in general, one should expect a nontrivial interplay between effects due to the quantum deformation and those due to spacetime curvature. For example, once the quantum deformation is taken into account the effects that are classically associated to space-time curvature acquire a new energy-dependence \cite{Marciano:2010gq,Amelino-Camelia:2012vzf, Rosati:2015pga, Barcaroli:2015eqe, Aschieri:2020yft}. Moreover, when space-time curvature is present, the description of the geometrical properties of momentum space is non-trivial 
\cite{Ballesteros:2021dob, Ballesteros:2017pdw, Ballesteros:2017kxj, Gutierrez-Sagredo:2019ipf} and it turns out that one needs to account for an enlarged momentum space, which includes additional coordinates associated to ``hyperbolic angular momentum''. In $3+1$ dimensions, the geometry of these momentum spaces is half of the 6+1
dimensional de Sitter space in the case of $\kappa$-dS, and half of a space with $SO(4, 4)$  invariance for  $\kappa$-AdS \cite{Ballesteros:2017pdw}.

We are not going to revisit the details of the derivation of the results we mentioned, since this would go beyond the scope of these notes. A thorough review and additional references  can be found in \cite{Ballesteros:2021dob}. 

Here we recall only some of the more interesting features concerning the interplay between the quantum deformation and curvature parameters, in order to  illustrate the previous remarks. The rotations sector is deformed into a quantum $so(3)$ algebra with deformation parameter given by $ \eta/\kappa= \sqrt{-\Lambda}/\kappa$:
\begin{align}
 \Delta  ( J_3  )&=   J_3 \otimes 1 +1 \otimes J_3 \,, \nonumber\\
\Delta  ( J_1  ) &=  J_1 \otimes e^{\frac{\eta}{\kappa} J_3} +1 \otimes J_1 \,,\qquad \Delta ( J_2 ) = J_2 \otimes e^{\frac{\eta}{\kappa}J_3}+1 \otimes J_2  \,, 
\label{deformedso3}
\end{align}
and whose deformed brackets read\footnote{Here we are using Poisson brackets instead of commutators because we are taking the semiclassical limit which turns a Hops algebra into a Poisson-Lie algebra.}
\be
\begin{array}{lll}
\multicolumn{3}{l}
{\displaystyle {\left\{ J_1,J_2 \right\}= \frac{e^{2\frac{\eta}{\kappa} J_3}-1}{2 {\eta}/{\kappa} } - \frac{{\eta}}{2 \kappa} \left(J_1^2+J_2^2\right)  \,,\qquad
\left\{ J_1,J_3 \right\}=-J_2 \,,\qquad
\left\{ J_2,J_3 \right\}=J_1 \,  . } }
 \end{array}
\label{be}
\ee

The translations sector, that provides the deformed composition law for momenta in the corresponding DSR model, as seen for the $\kappa$-Poincar\'e case in the previous section,  reads
\begin{align}
\Delta ( P_0 ) &=  P_0 \otimes 1+1 \otimes P_0  \, , \nonumber \\
\Delta ( P_1 )  &=  P_1\otimes   \cosh (\eta J_3 /\kappa) +e^{-P_0/\kappa}  \otimes P_1 
-  {\eta} K_2 \otimes     { \sinh (\eta J_3 /\kappa) }  \nonumber \\
&- \frac{\eta}{\kappa}P_3  \otimes J_1 + \frac{\eta^2}{\kappa} K_3 \otimes J_2  + \frac{\eta^2}{\kappa^2} \left( {\eta }K_1-P_2 \right)  \otimes J_1 J_2 e^{-\frac{\eta}{\kappa}J_3}   \nonumber\\
&-  \frac{\eta^2}{\kappa^2}  \left(  {\eta}K_2+P_1   \right)  \otimes      \left( J_1^2-J_2^2 \right)    e^{-\frac{\eta}{\kappa}J_3}  \, , \nonumber \\
\Delta ( P_2 ) &=  P_2\otimes  \cosh( \eta J_3 /\kappa) +e^{-P_0/\kappa}  \otimes P_2
+  \eta K_1 \otimes    { \sinh (\eta J_3 /\kappa) }  \nonumber \\
&- \frac{\eta}{\kappa} P_3  \otimes J_2 - \frac{\eta^2}{\kappa} K_3 \otimes J_1  - \frac{\eta^2}{\kappa^2} \left( \eta K_2+P_1 \right)  \otimes J_1 J_2 e^{-\frac{\eta}{\kappa}J_3}    \\
& - \frac 12 \frac{\eta^2}{\kappa^2}    \left(    \eta K_1- P_2 \right)  \otimes     \left( J_1^2-J_2^2 \right)       e^{-\frac{\eta}{\kappa}J_3}  \,, \nonumber \\
\Delta ( P_3 )  &=  P_3  \otimes 1 +e^{-P_0/\kappa}  \otimes P_3  + \frac{1}{\kappa}  \left(  \eta^2  K_2+ \eta  P_1 \right)  \otimes J_1  e^{-\frac{\eta}{\kappa}J_3}  \nonumber \\
& -\frac{1}{\kappa}     \left(  \eta^2  K_1- \eta P_2 \right) \otimes  J_2 e^{-\frac{\eta}{\kappa}J_3}  \, . \nonumber 
\end{align}
Notice that the deformed composition law for momenta involves the full Lorentz sector, which indicates that the construction of the  associated momentum space needs to include the Lorentz sector, as we discussed at the beginning of this subsection.
Moreover, the deformed brackets describe both non-commutativity due to space-time curvature ($\eta\neq 0$) and quantum deformation:
  \begin{align}
& \left\{ P_1, P_2 \right\}=- \eta^2\, \frac{  \sinh(2 \frac{\eta}{\kappa}J_3)}{2{\eta}/{\kappa} }- \frac{\eta}{2 \kappa}  \left( 2 P_3^2+ {\eta^2} (J_1^2+J_2^2)  \right)- \frac{\eta^5}{4 \kappa^3} e^{-2\frac{\eta}{\kappa}J_3} \left(J_1^2+J_2^2 \right)^2  \, , \nonumber \\
& \left\{ P_1, P_3 \right\}=\frac12 {\eta^2} J_2 \left( 1+e^{-2\frac{\eta}{\kappa}J_3} \left[1+   \frac{\eta^2}{\kappa^2}   \left( J_1^2 +J_2^2 \right) \right]  \right) + \frac{\eta}{\kappa}  P_2 P_3  \, , \\
& \left\{ P_2, P_3 \right\}=-\frac12 {\eta^2} J_1 \left( 1+e^{-2\frac{\eta}{\kappa}J_3} \left[1+  \frac{\eta^2}{\kappa^2}   \left( J_1^2 +J_2^2 \right) \right]  \right)  - \frac{\eta}{\kappa}   P_1 P_3\,. \nonumber
\end{align}
Finally, the Casimir of the algebra reads
\begin{align}
{\cal C}_{\kappa,\eta}&=2{\kappa^2}\left[ \cosh (P_0/\kappa)\cosh(\frac{\eta}{\kappa} J_3)-1 \right]+ {\eta^2} \cosh(P_0/\kappa) (J_1^2+
J_2^2) e^{-\frac{\eta}{\kappa}J_3} \nonumber\\
& -e^{P_0/\kappa} \left( \mathbf{P}^2 + {\eta^2} \mathbf{K}^2 \right)   \left[ \cosh(\frac{\eta}{\kappa}J_3)+ \frac {\eta^2 }{2\kappa^2}  (J_1^2+J_2^2)  e^{-\frac{\eta}{\kappa}J_3} \right]\nonumber\\
&+2 {\eta^2} e^{P_0/\kappa}  \left[ \frac{\sinh(\frac{\eta}{\kappa}J_3)}{\eta}\mathcal{R}_3+ \frac{1}{\kappa} \left( J_1\mathcal{R}_1 +J_2 \mathcal{R}_2+  \frac {\eta}{2\kappa} (J_1^2+J_2^2) \mathcal{R}_3 \right)  e^{-\frac{\eta}{\kappa} J_3} \right],
\end{align}
where $\mathcal{R}_a=\epsilon_{abc} K_b P_c $. As expected, in the $\kappa\to \infty$ limit we obtain the de Sitter Casimir, and in the $\eta\to 0$ limit, we obtain the $\kappa$-Poincar\'e  Casimir in the bicrossproduct basis~\eqref{eq:kPCasimir}.

A derivation of the DSR model that would correspond to the $\kappa$-(A)dS algebra in $3+1$ dimensions is still missing, due to the difficulty of defining the phase space of particles when coordinates and momenta are intertwined in such a way. Preliminary studies on the propagation of free particles in $1+1$ dimensions were recently performed \cite{Barcaroli:2015eqe}, see also Section \ref{sec:pheno}.

\section{Relative locality}
\label{sec:relloc}

We have seen in the previous sections that DSR models are most naturally described in momentum space rather than in spacetime. Indeed, a commonly accepted view, also supported  by the results concerning the emergence of DSR within more fundamental quantum gravity theories discussed in Section \ref{sec:fundamental}, is that DSR may characterize a semi-classical and ``non-gravitational'' regime of quantum gravity. That is, heuristically, the limit in which Newton and Planck constants are negligible, $G_N\to 0$ and $\hbar \to 0$, so that both quantum and gravitational effects are small, but their ratio $\hbar/G_N$ remains constant \cite{Amelino-Camelia:2011lvm, Amelino-Camelia:2011hjg}.  
In this regime, which is labeled the ``relative-locality regime'', for reasons that will be made clear in the following, modifications of standard physics governed by the Planck energy $E_P\sim\sqrt{\hbar/G_N}\neq 0$ can be present. Given the presence of this energy scale, it is natural to take as fundamental notion  that of momentum space,  assumed to be a pseudo-Riemannian manifold with an origin, a metric $g_{\mu\nu}(p)$, and a connection $\Gamma_\mu^{\nu\sigma}$, which can have torsion and non-metricity. The energy scale is then linked to the curvature scale of such manifold.

On the other hand, in this limit the Planck length $L_P\sim\sqrt{\hbar G_N}$ goes to zero, which smooths out possible small-scale properties of spacetime, such as noncommutativity. Nevertheless, this does not mean that we can describe spacetime as we usually do in general relativity or in quantum physics. In particular, the notion of locality becomes observer-dependent: the fact that two events take place at the same space-time point can only be established by observers close to the events themselves. The introduction of an observer-independent energy scale in DSR implies relativity of locality just as  the introduction of an observer-independent speed scale in special relativity implies relativity of simultaneity.

The relative locality proposal \cite{Amelino-Camelia:2011lvm, Amelino-Camelia:2011hjg} resulted from a deepening in
the understanding of the fate of the locality principle in DSR models \cite{Hossenfelder:2010tm, Smolin:2011izc, Amelino-Camelia:2010wxx, Jacob:2010vr, Amelino-Camelia:2011ebd}. Besides clarifying this important issue, the proposal also provided a more physical framework to understand the interpretation of the group manifold underlying the $\kappa$-Poincar\'e Hopf algebra  as a curved momentum space, which was discussed in the previous section \cite{Gubitosi:2011hgc, Amelino-Camelia:2011uwb, Gubitosi:2019ymi}.  In fact,  the relative-locality framework provides an interpretation based on the geometry of momentum space for deformations of on-shellness and of conservation laws of energy-momentum: the metric on momentum space is linked to the on-shell relation while the affine connection on momentum space is related to the law of composition of momenta, which enters into the  laws of conservation of energy-momentum.

\subsection{Geometry of momentum space}
\label{sec:momspacegeom}

One-particle system measurements allow the observer to determine the metric of momentum space through the dispersion relation, linked to the square of the geodesic distance from the origin to a point $p$ in momentum space corresponding to the momentum of the particle:
\begin{equation}
m^2= D^2(0,p) =\int_0^1 ds \sqrt{g_{\mu\nu}\dot p^\mu\dot p^\nu}\,,\label{eq:distance}
\end{equation}
where $p^\mu$ is the solution of the geodesic equation for the metric $g_{\mu\nu}$ \cite{Amelino-Camelia:2011uwb}.

In order to allow us to construct a relativistic model for particle kinematics, the metric $g_{\mu\nu}$ must be maximally symmetric, leading to only three options: Minkowski, de Sitter, or anti-de Sitter metrics. In most of the models studied so far, the metric is that of de Sitter \cite{Gubitosi:2011hgc, Amelino-Camelia:2011uwb, Amelino-Camelia:2013sba}. Anti-de Sitter momentum spaces have been explored \cite{ Arzano:2014jua, Lobo:2016blj}, but it is not clear whether they lead to viable models.

As we mentioned above, the momentum space manifold can have a non-metric connection, so that its geometry is not  determined completely by the metric. Specification of the addition rules of energy-momentum in particles interactions can be used to define the connection of the momentum space. This connection is in general non-metrical, in the sense that it is not the Levi-Civita connection given by the metric defined by the deformed dispersion relation. 

Considering the momenta of two particles, the original proposal of \cite{Amelino-Camelia:2011lvm, Amelino-Camelia:2011hjg} relies on the parallel transport of the momentum of one particle to the point in momentum space corresponding to the momentum of the other particle \cite{Amelino-Camelia:2013sba}, and defines the connection via
\be
\Gamma^{\tau \lambda}_\nu (k)\,=\,-\left.\frac{\partial^2  (p\oplus_{k}q)_\nu}{\partial p_\tau \partial q_\lambda}\right\rvert_{p=q=k}\,,
\label{k-connection}
\ee
where 
\be
(p\oplus_k q) \,\doteq\, k\oplus\left((\ominus{k}\oplus p)\oplus(\ominus{k}\oplus q)\right)\,,
\label{k-DCL}
\ee
and $\ominus$ is the so-called antipode operation of $\oplus$, such that $(\ominus p)\oplus p=(\oplus p)\ominus p=0$. 
The antisymmetric part of this connection is the torsion, which is linked to the noncommutativity of the composition law
\be
T^{\tau \lambda}_\nu (k)\,=\,-\left.\frac{\partial^2  \left((p\oplus_k q)-(q\oplus_k p)\right)_\nu}{\partial p_\tau \partial q_\lambda}\right\rvert_{p=q=k}\,,
\ee
while non-associativity of the composition law determines the connection curvature
\be
R^{\mu\nu\rho}_\sigma (k)\,=\,2 \left.\frac{\partial^3  \left((p\oplus_k q)\oplus_k r-p\oplus_k (q\oplus_k r)\right)_\sigma}{\partial p_{[\mu} \partial q_{\nu]} \partial r_\rho}\right\rvert_{p=q=r=k}\,,
\ee
where the bracket denotes the anti-symmetrization. In \cite{Amelino-Camelia:2011lvm, Freidel:2011mt}, the nonmetricity, defined from the metric and the connection as 
\be
N^{\mu\nu\rho}=\nabla^\rho g^{\mu\nu}(k)\,,
\ee
is claimed to be responsible for the leading order time-delay effect in the arrival of photons from distant sources, see also Section \ref{sec:pheno}.

Notice that because of the derivatives of the addition law appearing in the definition of  the torsion, one may have non-commutative composition laws that still produce a symmetric (i.e. torsionless) connection. This can be traced back to the fact that, when going beyond the first order in the deformation, the definition \eqref{k-connection} does not allow for a unique identification of the composition law based on a given connection (and in particular, symmetric connections can be associated to non-commutative composition laws). Motivated by this, an alternative proposal for the connection was provided in \cite{Amelino-Camelia:2013sba}. In this alternative proposal, there is a one-to-one correspondence between the composition law and the connection (at least up to second order in the deformation), and symmetric connections correspond to commutative composition laws. The drawback is that not all possible composition laws can be mapped to a connection, but only those satisfying a cyclic property \cite{Amelino-Camelia:2013sba}. The two proposals for the connection are equivalent when applied to the $\kappa$-Poincar\'e kinematics described in section \ref{sec:kPpheno}. {\bf Exercise:} Compute the connection \eqref{k-connection} associated to the composition law of  the $\kappa$-Poincar\'e model, eq. \eqref{eq:kPconslaw}. Use \cite{Gubitosi:2011hgc, Amelino-Camelia:2013sba} as guidance.

The reason why a geometrical description of the relativistic kinematics is useful is that it may allow to characterize the (DSR-)relativistic invariance of kinematics on momentum space in terms of constraints on the geometry. We have already discussed one such example in the case of the dispersion relation and the momentum space metric. Concerning the connection, a thorough analysis is provided in \cite{Amelino-Camelia:2013sba}.

\subsection{Spacetime and relativity of locality}
\label{sec:rellocspacetime}

Some of the most relevant phenomenological applications of the deformed kinematics encoded in the relative locality proposal require that  a notion of spacetime is provided. This is particularly important for studies of the time of flight, where one looks for a difference in the arrival time of particles with different energies emitted simultaneously by the same astrophysical source (see Section \ref{sec:pheno}). 

The relative locality framework provides a proposal for a description of spacetime  that is compatible with the deformed relativistic symmetries of the momentum space, allows for a description of the phase space of a single free particle, and allows us to describe the more complex case of interacting particles. Notice that this last point is especially nontrivial. In fact, when the momentum space is curved, one can take the momentum space as the base manifold and construct spacetime as the cotangent space to the momentum space at a given point $p$.\footnote{This is a completely analogous construction to the one of general relativity where momentum space is the cotangent space of the space-time manifold at a point in spacetime.} Such notion is well defined in the case of one free particle, because the particle lives on one point of the curved momentum manifold, $p$. Then one can define the free particle dynamics in a canonical way, with the role of spacetime and momentum space exchanged with respect to the usual construction: space-time coordinates $x^\mu$ are canonically conjugated to momenta via Poisson brackets,\footnote{See \cite{AmelinoCamelia:2011nt} for alternative, but physically equivalent, prescriptions.}
\be
\{x^u,p_\nu\} = \delta^\mu_\nu\,,
\label{eq:pb}
\ee
and the dynamics  of a free particle is  described  by the  action
\be
\label{eq:free_action}
S^{\text{free}} = \int \D \lambda \left( -x^\mu \dot p_\mu + \mathcal N \left( D(p)^2 - m^2\right) \right)\,.
\ee
The over-dot indicates the derivative with respect to the affine parameter $\lambda$ and the parameter $\mathcal N$ is a Lagrange multiplier enforcing  on-shellness $D(p)^2-m^2=0$ (see Eq. \eqref{eq:distance}).
Variation of \eqref{eq:free_action} with respect to $x^\mu$ and $p_\mu$ yields, respectively, conservation of momentum
\begin{align}
\dot p_\mu = 0\,,
\end{align}
and the evolution equation for the space-time coordinates:
\begin{align}
\label{eqn:xdot}
\dot x^\mu = - \mathcal N \frac{\partial \mathcal C}{\partial p_\mu}\,,
\end{align}
where $ \mathcal C(p) \equiv D(p)^2 - m^2$.

{\bf Exercise:} Show that the on-shell relation and the constraint equations in the case of the $\kappa$-Poincar\'e model are (use \cite{Gubitosi:2011hgc, Gubitosi:2019ymi} for guidance):
\begin{align}
m &= \kappa \,\arccosh \left( \cosh\frac{p_0}{\kappa}-e^{\frac{p_0}{\kappa}}\frac{\vec p^2}{2\kappa^2}\right)\,, \label{eq:rel_loc_disp_rel}\\
\dot p_\mu &= 0\,, \label{eq:rel_loc_const_p} \\
\frac{\partial x^i}{\partial x^0} &\equiv \frac{\dot x^i}{\dot x^0} = \frac{2\kappa p_i}{\kappa^2\left(e^{-2 p_0/\kappa}-1\right) + \vec p^2}\,. \label{eq:rel_loc_worldline}
\end{align}

In the case of the 1+1 dimensional $\kappa$-Poincar\'e model, after integrating the coordinates evolution  \eqref{eq:rel_loc_worldline} and using the on-shell relation \eqref{eq:rel_loc_disp_rel}, one finds the worldline \cite{Gubitosi:2011hgc}:
\be
x^{1}(x^{0}) = x^{1}(0) + v(p)x^{0}\,,\qquad v(p) =\frac{e^{p_0/\kappa}\sqrt{e^{2 p_0/\kappa}+1-2\,  e^{p_0/\kappa}\cosh(m/\kappa)}}{1-e^{p_0/\kappa}\cosh(m/\kappa)}\,. \label{eq:rel_loc_worldline1}
\ee
Because the momentum $p_\mu$ is a constant of motion,  the velocity $v(p)$ is constant as well. The linearity of the worldlines with respect to space-time coordinates indicates that spacetime is flat. The deformed expression for $v(p)$ can be attributed to the non-trivial geometry of momentum space.

When several interacting particles are considered, it is not obvious what momentum to use in order to build the spacetime at the interaction event. For each particle $I$ with momentum $p^I$, the construction outlined above requires a different set of space-time coordinates $x_I^\mu$, each living on the cotangent space of the momentum manifold at a different point $p^I$ and canonically conjugate to the associated momentum. If the particles are not interacting, the total action is given by the sum of the free actions of each particle:
 \begin{align}
S^{tot}&=\sum_I S_I^{\text{free}}\,, \nonumber \\
S_I^{\text{free}} &= \int_{-\infty}^{\infty} \D \lambda \left( -x_I^\mu \dot p^I_\mu + \mathcal N_I \left( D(p^I)^2 - m_I^2\right) \right)\,. \label{eq:free_action_multiparticle}
\end{align}

 If the particles are interacting, it does not make sense to ask that the coordinates $x_I^\mu$ take the same value for all $I$'s at the interaction event, because  the space-time coordinates of each particle live in different cotangent spaces. The solution provided within the relative locality framework  is to introduce a boundary interaction term in the action, with a constraint that enforces momentum conservation at the interaction \cite{AmelinoCamelia:2011bm}. In
the case of a single vertex (interaction among $n$ incoming and $m$ outgoing particles) the total action is:
 \begin{align}
S^{tot}&=\sum_{I=1}^{n+m} S_I^{\text{free}}+S^{int} \,, \nonumber \\
S_I^{\text{free}} &= \pm\int_{\lambda_{0}^{I}}^{\pm \infty} \D \lambda \left( -x_I^\mu \dot p^I_\mu + \mathcal N_I \left( D(p^I)^2 - m_I^2\right) \right)\,, \nonumber\\
S^\text{int} &= z^\mu \mathcal K_\mu(p_1(\lambda_0^{1}),\dots,p_n(\lambda_0^{n}), p_{n+1}(\lambda_0^{n+1}),\dots,p_m(\lambda_0^{m}))\,, \label{eq:rel_loc_action_total}
\end{align}
where the $\pm$ sign is chosen according to whether the $I$-th particle is outgoing or incoming, $\lambda_0^{I}$ is the value of the affine parameter at the endpoint of the worldline of each particle where the interaction occurs, and $z^\mu$ is a Lagrange multiplier enforcing the conservation law  $\mathcal K_\mu(p_1(\lambda_0^{1}),\dots,p_n(\lambda_0^{n}), p_{n+1}(\lambda_0^{n+1}),\dots,p_m(\lambda_0^{m}))=0$. $K_\mu$ accounts  for the deformed composition of momenta (for details see \cite{AmelinoCamelia:2011nt, Gubitosi:2019ymi})
\be
p_1\oplus \dots \oplus p_n= p_{n+1}\oplus \dots \oplus p_m\,. \label{eq:conservation_law}
\ee

 Upon varying the action one gets  similar constraints for each interacting particle as those found for the free particle, Eqs. \eqref{eq:rel_loc_disp_rel}-\eqref{eq:rel_loc_worldline}. Additionally, the interaction term yields an additional constraint on the endpoints of the worldlines at the interaction,
\be 
x^\mu_I(\lambda_0^{I}) = \mp z^\nu \frac{\partial \mathcal K_\nu }{\partial p^I_\mu }\bigg|_{\lambda=\lambda_0}\,,\label{eq:rel_loc_constraint_interaction}
\ee
where the upper (lower) sign is for outgoing (incoming) particles.\footnote{This result was recently rederived using a line element in phase space for a multi-particle system in~\cite{Relancio:2021ahm}} In the case of special relativity, $\mathcal K_\mu = p_1+ \dots + p_n - (p_{n+1}+ \dots + p_m)$,  and all the worldlines simply end up at the interaction point $x^\mu_I=z^\mu$, so that the interaction is local. If the nonlinearity of momentum space induces nonlinear corrections to the composition law of momenta (as e.g. in \eqref{eq:kPconslaw}), then the worldlines will have in general different endpoints, since $ \frac{\partial \mathcal K_\nu }{\partial p^I_\mu }\neq  \frac{\partial \mathcal K_\nu }{\partial p^J_\mu }$. In this case, only if the interaction happens at $z^\mu=0$ then all worldlines end at $x^\mu_I=z^\mu=0$. If instead $z^\mu\neq 0$, then each worldline  ends at a different value of $x^\mu_I$. This is a manifestation of relative locality: only a local observer, $z^\mu=0$, sees the interaction as local, while other observers, $z^\mu\neq 0$, see each worldline ending at a slightly different point. A more in-depth discussion can be found in \cite{Amelino-Camelia:2011uwb}, Section V.

It can be shown that this space-time picture is compatible with the deformed relativistic symmetries, see \cite{Gubitosi:2019ymi, Amelino-Camelia:2011uwb}. In particular, the worldlines transform covariantly under translations and boosts, if the corresponding generators are taken to be the ``total generators'' in the sense already discussed in Section \ref{sec:kPpheno}, see Eq.\eqref{eq:totalboost}. Moreover, the ``interaction coordinates'' $z^\mu$ transform  as the space-time coordinates of a single particle with momentum given by the total momentum of the vertex.

Notice that the action \eqref{eq:rel_loc_action_total} can be further expanded to include several interaction vertices, sharing some of the particles involved.
In doing so, however, one runs into what is known as ``history problem'' (sometimes also called ``spectator problem''). Because the action is invariant under the action of ``total generators'', one needs to know the whole sequence of causally connected vertices in order to correctly define such operators and  transform any individual vertex \cite{Gubitosi:2019ymi}. Understanding how to solve this issue is still an open problem  currently under  study.

\section{Deformed kinematics on curved momentum space}
\label{sec:comparison}

In the previous section we have discussed a possible way to describe a curved momentum space which takes into account a relativistic deformed kinematics. 
In that approach, the starting ingredients are the free particle energy-momentum dispersion relation and the addition law of momenta in interactions. From these, one can derive the geometrical properties of the momentum space (metric and connection) following the prescriptions discussed in \ref{sec:momspacegeom}.
In this section we develop a different perspective~\cite{Carmona:2019fwf}: we start from the geometry of a maximally symmetric curved momentum space and derive all the ingredients of a deformed kinematics, preserving a relativity principle~\cite{Amelino-Camelia:2011gae, Carmona:2012un, Amelino-Camelia:2013sba,Carmona:2016obd}.

\subsection{Definition of the deformed kinematics}

A relativistic deformed kinematics (we will see in Sec.~\ref{sec:diagram}  that this construction preserves a relativity principle) can be obtained by identifying the isometries of the momentum space metric with the composition law and the Lorentz transformations in the one-particle system, fixing the dispersion relation.

It is well known that in a four dimensional maximally symmetric space, there are 10 isometries~\cite{Weinberg:1972kfs}. An isometry is a transformation $k\to k'$   that, when acting on a momentum metric   $g_{\mu\nu}(k)$, does not change the form of the metric, i.e.,
\be
  g_{\mu\nu}(k')=\frac{\partial k'_\mu}{\partial k_\rho} \frac{\partial k'_\nu}{\partial k_\sigma} g_{\rho\sigma}(k)\, .
\ee

By choosing a system of coordinates   such   that $g_{\mu\nu}(0)=\eta_{\mu\nu}$,   we can write the isometries as
\be
k'_\mu=[T_a(k)]_\mu=T_\mu(a, k)\,, \quad\quad\quad k'_\mu=[J_\omega(k)]_\mu \,=\,J_\mu(\omega, k)\,,
\ee
where $a$ is a set of four   parameters and $\omega$ of six, and 
\be
T_\mu(a, 0)=a_\mu\,, \quad\quad\quad J_\mu(\omega, 0)=0\,.
\ee
Here $J_\mu(\omega, k)$ are the 6 Lorentz isometries (three rotations and three boosts) which form a subgroup (Lorentz algebra), leaving the origin in momentum space invariant. On the other hand, $T_\mu(a, k)$ are the other 4 isometries associated to translations which transform the origin. This idea was also considered in~\cite{Lobo:2016blj} but, as we will see, there is some arbitrariness that needs to be fixed in order to obtain the desired kinematics.

Therefore, the isometries $k'_\mu = J_\mu(\omega, k)$ are the Lorentz transformations of the one-particle system, being $\omega$ the six parameters of a Lorentz transformation. In order to define the dispersion relation $\mathcal{C}_{\kappa}(k)$, we can use any arbitrary function of the distance from the origin to a point $k$, in such a way that special relativity is recovered  when taking the limit in which the high-energy scale tends to infinity.\footnote{Note that in the previous section the squared distance was identified with the squared of the distance in momentum space, but, as discussed after Eq.~\eqref{eq:kPCasimir}, any function of the Casimir will be also a Casimir.} Since the distance is invariant under a Lorentz transformation, the equality $\mathcal{C}_{\kappa}(k)=\mathcal{C}_{\kappa}(k')$ holds, which allows us to determine the Casimir directly from $J_\mu(\omega, k)$ without computing the explicit form of the distance:
\be
\frac{\partial \mathcal{C}_{\kappa}(k)}{\partial k_\mu} \,{\cal J}^{\alpha\beta}_\mu(k)=0 \,.
\label{eq:casimir_J}
\ee

The other 4 isometries $k'_\mu = T_\mu(a, k)$ are related to  translations in momentum space and define the composition law $p\oplus q$ of two momenta $p$, $q$ through
\be
(p\oplus q)_\mu \,\doteq\, T_\mu(p, q)\,.
\label{DCL-translations}
\ee
Indeed one can see that this composition is related with the composition of translations
\be
p\oplus q\,=\,T_p(q)\,=\,T_p(T_q(0))\,=\,(T_p \circ T_q)(0)\,.
\label{T-composition}
\ee
In the following, we will discuss the possible different definitions of translations for a given metric.

Then, the deformed kinematics can be obtaining from a momentum metric via
\be
\begin{split}
g_{\mu\nu}(T_a(k))=&\frac{\partial T_\mu(a, k)}{\partial k_\rho} \frac{\partial T_\nu(a, k)}{\partial k_\sigma} g_{\rho\sigma}(k)\,,\\
g_{\mu\nu}(J_\omega(k)) \,=\,& \frac{\partial J_\mu(\omega, k)}{\partial k_\rho} \frac{\partial J_\nu(\omega, k)}{\partial k_\sigma} g_{\rho\sigma}(k)\,.
\end{split}
\label{T,J}
\ee
These equations must be satisfied for any $a$, $\omega$. One can see, from the limit $k\to 0$ in \eqref{T,J}
\be
\begin{split}
g_{\mu\nu}(a) \,=&\, \left[\lim_{k\to 0} \frac{\partial T_\mu(a, k)}{\partial k_\rho}\right] \, 
\left[\lim_{k\to 0} \frac{\partial T_\nu(a, k)}{\partial k_\sigma}\right] \,\eta_{\rho\sigma}\,,  \\
\eta_{\mu\nu} \,=&\, \left[\lim_{k\to 0} \frac{\partial J_\mu(\omega, k)}{\partial k_\rho}\right] \,  
\left[\lim_{k\to 0} \frac{\partial J_\nu(\omega, k)}{\partial k_\sigma}\right] \,\eta_{\rho\sigma}\,,
\end{split}
\ee
that
\be
\lim_{k\to 0} \frac{\partial T_\mu(a, k)}{\partial k_\rho}=\delta^\rho_\alpha e_\mu^\alpha(a)\,, \quad\quad\quad
\lim_{k\to 0} \frac{\partial J_\mu(\omega, k)}{\partial k_\rho}=L_\mu^\rho(\omega)\,,
\label{e,L}
\ee
where $e_\mu^\alpha(k)$ is the inverse of the tetrad of the momentum space metric,\footnote{Note that the metric $g_{\mu\nu}$ is the inverse of $g^{\mu\nu}$.} and $L_\mu^\rho(\omega)$ is the standard Lorentz transformation matrix with parameters $\omega$. From Eq.~\eqref{DCL-translations} and Eq.~\eqref{e,L}, one obtains
\be
\lim_{k\to 0} \frac{\partial(a\oplus k)_\mu}{\partial k_\rho}=\delta^\rho_\alpha e_\mu^\alpha(a)\,,
\label{magicformula}
\ee
which establishes a relationship between the composition law and the tetrad.

We can write, for infinitesimal transformations
\be
T_\mu(\epsilon, k) = k_\mu + \epsilon_\alpha {\cal T}_\mu^\alpha(k)\,, \quad\quad\quad
J_\mu(\epsilon, k) = k_\mu + \epsilon_{\beta\gamma} {\cal J}^{\beta\gamma}_\mu(k)\,,
\label{infinit_tr}
\ee
then Eq.~\eqref{T,J} becomes 
\be
\frac{\partial g_{\mu\nu}(k)}{\partial k_\rho} {\cal T}^\alpha_\rho(k)=\frac{\partial{\cal T}^\alpha_\mu(k)}{\partial k_\rho} g_{\rho\nu}(k) +
\frac{\partial{\cal T}^\alpha_\nu(k)}{\partial k_\rho} g_{\mu\rho}(k)\,,
\label{cal(T)}
\ee
\be
\frac{\partial g_{\mu\nu}(k)}{\partial k_\rho} {\cal J}^{\beta\gamma}_\rho(k) \,=\,
\frac{\partial{\cal J}^{\beta\gamma}_\mu(k)}{\partial k_\rho} g_{\rho\nu}(k) +
\frac{\partial{\cal J}^{\beta\gamma}_\nu(k)}{\partial k_\rho} g_{\mu\rho}(k)\,,
\label{cal(J)}
\ee
which define the Killing vectors ${\cal J}^{\beta\gamma}$, but do not completely determine ${\cal T}^\alpha$. This can be understood from the fact that if ${\cal T}^\alpha$, ${\cal J}^{\beta\gamma}$ are a solution of the Killing equations (\eqref{cal(T)}-\eqref{cal(J)}), then ${\cal T}^{\prime \alpha} = {\cal T}^\alpha + c^\alpha_{\beta\gamma} {\cal J}^{\beta\gamma}$ is also a solution of Eq.~(\ref{cal(T)}) for any arbitrary constants $c^\alpha_{\beta\gamma}$, and then,  $T'_\mu(\epsilon, 0)=T_\mu(\epsilon, 0)=\epsilon_\mu$,  where $T'_\mu(\epsilon,k)= k_\mu + \epsilon_\alpha {\cal T}_\mu^{\prime \alpha}(k)$. We can eliminate this ambiguity by taking into account that    the isometry generators close an algebra~\cite{Dubovsky:2006vk}. Therefore we can ask the isometry generators, written as  
\be
T^\alpha=x^\mu {\cal T}^\alpha_\mu(k), \quad\quad\quad J^{\alpha\beta}=x^\mu {\cal J}^{\alpha\beta}_\mu(k)\,,
\label{generators_withx}
\ee
which lead to the Poisson brackets
\begin{align}
  &\{T^\alpha, T^\beta\}=x^\rho \left(\frac{\partial{\cal T}^\alpha_\rho(k)}{\partial k_\sigma} {\cal T}^\beta_\sigma(k) - \frac{\partial{\cal T}^\beta_\rho(k)}{\partial k_\sigma} {\cal T}^\alpha_\sigma(k)\right)\,, \\
  &\{T^\alpha, J^{\beta\gamma}\}=x^\rho \left(\frac{\partial{\cal T}^\alpha_\rho(k)}{\partial k_\sigma} {\cal J}^{\beta\gamma}_\sigma(k) - \frac{\partial{\cal J}^{\beta\gamma}_\rho(k)}{\partial k_\sigma} {\cal T}^\alpha_\sigma(k)\right)\,,
\end{align}
to close a particular algebra.  Note that $x^\mu$ are canonically conjugated variables of $k_\nu$, satisfying the Poisson brackets of~\eqref{eq:pb}. This ambiguity in defining the translations is just the ambiguity in the choice of the isometry algebra, leading each choice to   a different composition law, and therefore to different relativistic deformed kinematics.   Note that the dispersion relation is univocally defined once the metric is given, while the composition law can take different forms depending on the choice of the generators of translations $T^\alpha$, leading to different kind of deformed kinematics (see~\cite{Carmona:2012un,Carmona:2016obd} for a systematic way of constructing deformed kinematics order by order in the high-energy scale), as we will see in the following.

\subsection{Relativistic deformed kinematics}
\label{sec:diagram}

In this part we demonstrate that the previously defined kinematics are in fact relativistic. This can be understood from the following diagram:
\begin{center}
\begin{tikzpicture}
\node (v1) at (-2,1) {$q$};
\node (v4) at (2,1) {$\bar q$};
\node (v2) at (-2,-1) {$p \oplus q$};
\node (v3) at (2,-1) {$(p \oplus q)^\prime$};
%\draw [->] (v1) edge (v4);
\draw [->] (v1) edge (v2);
\draw [->] (v4) edge (v3);
\draw [->] (v2) edge (v3);
%\node at (0,1.3) {${\cal J}_R$};
\node at (-2.6,0) {$T_p$};
\node at (2.7,0) {$T_{p^\prime}$};
\node at (0,-1.4) {$J_\omega$};
\end{tikzpicture}
\end{center}
where the momentum with prime denotes the transformation through ${\cal J}_\omega$, and $T_p$, $T_{p'}$ are the translations with parameters $p$ and $p'$, respectively. We can define $\bar{q}$ as a point in momentum space satisfying  
\be
(p\oplus q)'=(p' \oplus \bar{q})\,.
\label{qbar1}
\ee
From this definition it is easy to note that for $q=0$, also $\bar{q}=0$, and for any other value $q\neq 0$, the point $\bar{q}$ is obtained from $q$ by an isometry, being this a composition of  the translation $T_p$, a Lorentz transformation $J_\omega$, and the inverse of the translation $T_{p'}$. This transformation is obviously an isometry, due to the fact that the isometries are a group of transformations, and therefore, any composition of isometries is also an isometry. Since we have proved that there is an isometry from  $q\rightarrow \bar{q}$, leaving the origin invariant, then the distance of both points to the origin are the same, which is tantamount to say
\be
\mathcal C_\kappa(q)=\mathcal C_\kappa(\bar{q})\,.
\label{qbar2}
\ee
From Eqs.~\eqref{qbar1}-\eqref{qbar2} one can see that the deformed kinematics with ingredients $C$ and $\oplus$ is a relativistic deformed kinematics when identifying  the momenta $(p', \bar{q})$ as the two-particle Lorentz transformation of $(p, q)$. Indeed, Eq.~(\ref{qbar1}) implies that the composition law is invariant under the previously defined Lorentz transformation and Eq.~(\ref{qbar2}), together with $\mathcal C_\kappa(p)=\mathcal C_\kappa(p')$, that the deformed dispersion relation of both momenta is also Lorentz invariant. From this definition of the two-particle Lorentz transformations, one of the points ($p$) transforms as a single momentum, but for the other one ($q$) the transformation generally depends of both momenta (which indeed is the case of several examples obtained within Hopf algebras \cite{Gubitosi:2011hgc, Gubitosi:2019ymi, Arzano:2022ewc}).

%(which indeed is the case of several examples obtained within Hopf algebras~\cite{Gutt1983AnE,BALLESTEROS1995137,Majid:1995qg,Gracia-Bondia:2001ynb,Hammou:2001cc,Lukierski:2005fc,Amelino-Camelia:2011ycm,Amelino-Camelia:2017pne,Ciric:2017rnf,DimitrijevicCiric:2018blz,Novikov:2019kit,Lizzi:2021dud,Gubitosi:2021itz} and beyond~\cite{Meljanac:2009ej,Battisti:2010sr}), as we will see in the following for the particular example of $\kappa$-Poincaré. GG: I replaced these generic references to Hopf algebras to references where the two-particle transformation was actually discussed

Other geometrical approaches to deformed dispersion relations and  Lorentz transformations compatible with it at the one particle level were considered in the literature. In particular, one way to consider such modification of the special relativistic kinematics is through a velocity or momentum dependent spacetime, known as Finsler~\cite{zbMATH02613491}  and Hamilton~\cite{miron2001geometry} geometries, respectively. The case of Finsler geometries was considered in~\cite{Girelli:2006fw,Amelino-Camelia:2014rga,Letizia:2016lew}, while~\cite{Barcaroli:2016yrl,Barcaroli:2015xda,Barcaroli:2017gvg} were devoted to Hamilton spaces. In this geometrical constructions a clear connection with a deformed composition law and two-particle Lorentz transformations is still missing. 

\subsection{\texorpdfstring{$\kappa$}{k}-Poincaré relativistic kinematics}
\label{sec:examples}

Here we show how the kinematics of $\kappa$-Poincaré defined in section \ref{sec:kPpheno} can be obtained from the previous prescription, and explain how other models can also be defined in this context.  

Let us consider an isotropic kinematics, for which the general form of the algebra of the generators of isometries must be  
\be
\{T^0, T^i\}=\frac{c_1}{\kappa} T^i + \frac{c_2}{\kappa^2} J^{0i}, \quad\quad\quad \{T^i, T^j\}=\frac{c_2}{\kappa^2} J^{ij}\,,
\label{isoRDK}
\ee
where we impose the generators $J^{\alpha\beta}$ to satisfy the standard Lorentz algebra, and because of the fact that isometries are a group, the Poisson brackets of $T^\alpha$ and $J^{\beta\gamma}$ are fixed by Jacobi identities. Different algebras of the generators of translations, i.e., different choices  of the coefficients $(c_1/\kappa)$ and $(c_2/\kappa^2)$, will lead to different composition laws. 

For the simple case where $c_2=0$ in Eq.~\eqref{isoRDK},  the generators of translations close a subalgebra\footnote{We have reabsorbed the coefficient $c_1$ in the scale $\kappa$.}    
\be
\{T^0, T^i\}=\pm \frac{1}{\kappa} T^i\,.
\label{Talgebra}
\ee
A well known result of differential geometry (see Ch.6 of Ref.~\cite{Chern:1999jn}) is that, when the generators of left-translations $T^\alpha$ transforming $k \to T_a(k) = (a\oplus k)$ form a Lie algebra, also the generators of right-translations $\tilde{T}^\alpha$ transforming $k \to (k\oplus a)$, close the same algebra but with a different sign
\be
 \{\tilde{T}^0, \tilde{T}^i\}=\mp \frac{1}{\kappa} \tilde{T}^i \,.
\label{Ttildealgebra}
\ee   
We have found the explicit relation between the infinitesimal right-translations and the tetrad of the momentum metric in Eq.~\eqref{magicformula}, which gives
\be
(k\oplus\epsilon)_\mu=k_\mu+\epsilon_\alpha e^\alpha_\mu\equiv \tilde{T}_\mu(k,\epsilon)\,.
\ee
Comparing with Eq.~\eqref{infinit_tr} and Eq.~\eqref{generators_withx}, we see that right-translation generators are given by
\be
\tilde{T}^\alpha=x^\mu e^\alpha_\mu(k)\,.
\label{Ttilde}
\ee

Both algebras \eqref{Talgebra}-\eqref{Ttildealgebra} satisfy $\kappa$-Minkowski non-commutativity \eqref{kminkn}, so the problem of finding a tetrad $e^\alpha_\mu(k)$ fulfilling the algebra of Eq.~(\ref{Ttildealgebra}) is tantamount to obtaining a representation of this non-commutativity written in terms of canonical coordinates of the phase space. As a particular solution, one can see that the following choice of the tetrad
\be
e^0_0(k)=1\,, \quad\quad\quad e^0_i(k)=e^i_0(k)=0\,, \quad\quad\quad e^i_j (k)=\delta^i_j e^{\mp k_0/\kappa}\,,
\label{bicross-tetrad}
\ee
leads to a representation of $\kappa$-Minkowski non-commutativity. 

In order to obtain the finite translations $T_\mu(a,k)$, which form a subgroup inside the isometry group, Eq.~\eqref{e,L} can be  generalized to define a transformation that does not change the form of the tetrad:
\be
e_\mu^\alpha(T(a, k))=\frac{\partial T_\mu(a, k)}{\partial k_\nu} \,e_\nu^\alpha(k)\,.
\label{T(a,k)}
\ee
If $T_\mu(a,k)$ is a solution to the previous equation, leaving invariant the form of the tetrad, the metric will also be invariant, and it is then an isometry. These previously defined translations form a group,  since the composition of two transformations also leaves the tetrad invariant. Indeed, we can solve Eq. \eqref{T(a,k)}  for the tetrad in Eq.~\eqref{bicross-tetrad}, obtaining [{\textbf{Exercise}}]
\be
T_0(a, k)=a_0 + k_0, \quad\quad\quad T_i(a, k)=a_i + k_i e^{\mp a_0/\kappa}\,,
\ee
so the composition law is 
\be
(p\oplus q)_0=T_0(p, q)=p_0 + q_0\,, \quad\quad\quad
(p\oplus q)_i=T_i(p, q)=p_i + q_i e^{\mp p_0/\kappa}\,,
\label{kappa-DCL}
\ee
which is the one obtained  in the bicrossproduct basis of  $\kappa$-Poincaré kinematics~\eqref{coprodbicr} (up to a sign depending on the choice of the initial sign of $\kappa$ in Eq.~\eqref{bicross-tetrad}).

From~\eqref{eq:casimir_J} one can obtain the dispersion relation, where ${\cal J}^{\alpha\beta}$ are the infinitesimal Lorentz transformations satisfying Eq.~\eqref{cal(J)} with the metric $g_{\mu\nu}(k)=e^\alpha_\mu(k)\eta_{\alpha\beta}e^\beta_\nu(k)$ defined by the tetrad~\eqref{bicross-tetrad}:
\be
\begin{split}
&0=\frac{{\cal J}^{\alpha\beta}_0(k)}{\partial k_0}\,, \quad
0=- \frac{{\cal J}^{\alpha\beta}_0(k)}{\partial k_i} e^{\mp 2k_0/\kappa} + \frac{{\cal J}^{\alpha\beta}_i(k)}{\partial k_0}\,, \\
&\pm \frac{2}{\kappa} {\cal J}^{\alpha\beta}_0(k) \delta_{ij}=- \frac{\partial{\cal J}^{\alpha\beta}_i(k)}{\partial k_j} - \frac{\partial{\cal J}^{\alpha\beta}_j(k)}{\partial k_i}\,.
\end{split}
\ee
One  finally gets
 \be
{\cal J}^{0i}_0(k)=-k_i\,, \quad \quad \quad {\cal J}^{0i}_j(k)\,=\, \pm \delta^i_j \,\frac{\kappa}{2} \left[e^{\mp 2 k_0/\kappa} - 1 - \frac{\vec{k}^2}{\kappa^2}\right] \pm \,\frac{k_i k_j}{\kappa}\,,
\label{eq:j_momentum_space}
\ee 
which is equivalent to~\eqref{Lorentz transalg}, and so
\be
\mathcal{C}_{\kappa}(k)=\kappa^2 \left(e^{k_0/\kappa} + e^{-k_0/\kappa} - 2\right) - e^{\pm k_0/\kappa} \vec{k}^2  \,,
\label{eq:casimir_momentum_space}
\ee
which is the same function of the momentum that defines the dispersion relation of $\kappa$-Poincaré kinematics in the bicrossproduct basis~\eqref{eq:kPCasimir} (up to the sign in $\kappa$).

Finally, using the diagram of Sec.~\ref{sec:diagram}, we can find $\bar{q}$ satisfying
\be
(p\oplus q)'=p'\oplus \bar{q}\,.
\ee
Equating both expressions and taking only the linear terms in $\epsilon_{\alpha\beta}$ (parameters of the infinitesimal Lorentz transformation) one arrives to the equation
\be
\epsilon_{\alpha\beta} {\cal J}^{\alpha\beta}_\mu(p\oplus q)=\epsilon_{\alpha\beta} \frac{\partial(p\oplus q)_\mu}{\partial p_\nu} {\cal J}^{\alpha\beta}_\nu(p) + \frac{\partial(p\oplus q)_\mu}{\partial q_\nu} (\bar{q}_\nu - q_\nu)\,.
\ee
From the composition law of \eqref{kappa-DCL} with the minus sign, we find
\begin{align}
& \frac{\partial(p\oplus q)_0}{\partial p_0}=1\,, \quad
\frac{\partial(p\oplus q)_0}{\partial p_i}=0\,, \quad
\frac{\partial(p\oplus q)_i}{\partial p_0}=- \frac{q_i}{\kappa} e^{-p_0/\kappa}\,, \quad
\frac{\partial(p\oplus q)_i}{\partial p_j}=\delta_i^j\,, \\
& \frac{\partial(p\oplus q)_0}{\partial q_0}=1\,, \quad  \frac{\partial(p\oplus q)_0}{\partial q_i}=0\,, \quad
\frac{\partial(p\oplus q)_i}{\partial q_0}=0\,, \quad  \frac{\partial(p\oplus q)_i}{\partial q_j}=\delta_i^j e^{-p_0/\kappa}\,.
\end{align}
Therefore, we obtain
\be
\begin{split}
\bar{q}_0 \,&=\, q_0 + \epsilon_{\alpha\beta} \left[{\cal J}^{\alpha\beta}_0(p\oplus q) - {\cal J}^{\alpha\beta}_0(p)\right]\,, \\
\bar{q}_i \,&=\, q_i + \epsilon_{\alpha\beta} \, e^{p_0/\kappa} \, \left[{\cal J}^{\alpha\beta}_i(p\oplus q) - {\cal J}^{\alpha\beta}_i(p) + \frac{q_i}{\kappa} e^{-p_0/\kappa} {\cal J}^{\alpha\beta}_0(p)\right]\,,
\end{split}
\label{eq:jr_momentum_space}
\ee
and one can check that this is the Lorentz transformation of the two-particle system of $\kappa$-Poincaré in the bicrossproduct basis~\eqref{eq:kpboostcoproduct}.

For the choice of the tetrad in Eq.~\eqref{bicross-tetrad}, the metric in momentum space reads\footnote{This is the de Sitter metric written in the comoving coordinate system used in Ref.~\cite{Gubitosi:2011hgc,Carmona:2019fwf}.}
\be
g_{00}(k)=1\,, \quad\quad\quad g_{0i}(k)=g_{i0}(k)=0\,, \quad\quad\quad g_{ij}(k)=- \delta_{ij} e^{\mp 2k_0/\kappa}\,,
\label{bicross-metric}
\ee
which is a de Sitter momentum space with curvature $(12/\kappa^2)$ [{\textbf{Exercise}}].

This shows that the $\kappa$-Poincaré kinematics in the bicrossproduct basis~\cite{KowalskiGlikman:2002we} can be completely obtained from the geometric ingredients of a de Sitter momentum space with the choice of the tetrad of Eq.~\eqref{bicross-tetrad}. By using different choices of tetrad, such that the generators of Eq.~\eqref{Ttilde} close the algebra Eq.~\eqref{Ttildealgebra}, one can find the $\kappa$-Poincaré kinematics in different bases. Therefore, the different bases of $\kappa$-Poincaré can be geometrically interpreted as  different choices of coordinates in de Sitter space. 

Different relativistic kinematics, outside the Hopf algebra scheme, can be obtained in the aforementioned framework, such as Snyder kinematics, which is a very particular example from the point of view of Lorentz symmetry. Indeed, it is compatible with linear Lorentz invariance in both one- and two-particle systems. The deformed addition law is given by~\cite{Battisti:2010sr}
\begin{align}
(p\oplus q)_\mu &= p_\mu  \left(\sqrt{1+\frac{q^2}{\Lambda^2}}+\frac{p_{\rho} \eta^{\rho\nu} q_{\nu} }{\Lambda^2\left(1+\sqrt{1+p^2/\Lambda^2}\right)}\right)+ q_\mu \,.
\end{align}
{\textbf{Exercise:}} Show that Snyder kinematics can be derived imposing  $c_1=0$ in Eq.~\eqref{isoRDK}. 

Moreover, the kinematics known as hybrid models~\cite{Meljanac:2009ej} can be obtained when $c_1$, $c_2$ are non-zero. As a final note, it is important to notice that, with the construction discussed here, different kinematics (with different composition laws) are related to the same metric, and therefore, also to the same dispersion relation.

\subsubsection{Comparison with previous works}
\label{relative_locality_comparison}

In this section, we will compare the prescription followed in this section with the one proposed in Ref.~\cite{AmelinoCamelia:2011bm}. This comparison can only be carried out for the $\kappa$-Poincaré kinematics, since as we will see, the associativity property of the composition law plays a crucial role. In order to make the comparison, we can derive with respect to  $p_\tau$  the equation of the invariance of the tetrad under translations Eq.~\eqref{T(a,k)}, written in terms of the composition law
\be
\frac{\partial e^\alpha_\nu(p\oplus q)}{\partial p_\tau}=\frac{\partial  e^\alpha_\nu(p\oplus q)}{\partial (p\oplus q)_\sigma} \frac{\partial(p\oplus q)_\sigma}{\partial p_\tau}=\frac{\partial^2(p\oplus q)_\nu}{\partial p_\tau \partial q_\rho} e^\alpha_\rho(q)\,.
\ee 
One can find the second derivative of the composition law
\be
\frac{\partial^2(p\oplus q)_\nu}{\partial p_\tau \partial q_\rho}=e^\rho_\alpha(q) \frac{\partial  e^\alpha_\nu(p\oplus q)}{\partial (p\oplus q)_\sigma} \frac{\partial(p\oplus q)_\sigma}{\partial p_\tau}\,,
\ee
where $e^\nu_\alpha$ is the inverse of $e^\alpha_\nu$, $e^\alpha_\nu e^\mu_\alpha=\delta^\mu_\nu$. 
But also using Eq.~\eqref{T(a,k)}, one has
\be
e^\rho_\alpha(q)=\frac{\partial(p\oplus q)_\mu}{\partial q_\rho} e^\mu_\alpha(p\oplus q)\,,
\label{magicformula2}
\ee
and then 
\be
\frac{\partial^2(p\oplus q)_\nu}{\partial p_\tau \partial q_\rho} + \Gamma^{\sigma\mu}_\nu(p\oplus q) \,\frac{\partial(p\oplus q)_\sigma}{\partial p_\tau} \,\frac{\partial(p\oplus q)_\mu}{\partial q_\rho}=0\,,
\label{geodesic_tetrad}
\ee
where 
\be
\Gamma^{\sigma\mu}_\nu(k) \,\doteq\, - e^\mu_\alpha(k) \, \frac{\partial  e^\alpha_\nu(k)}{\partial k_\sigma}\,.
\label{e-connection}
\ee
It can be checked that the combination of tetrads and derivatives appearing in Eq.~\eqref{e-connection} in fact transforms like a connection~\cite{Weinberg:1972kfs} under a change of momentum coordinates [{\textbf{Exercise}}].

In Ref.~\cite{Amelino-Camelia:2013sba}, it is proposed another way to define a connection and a composition law in momentum space through parallel transport, establishing a link between these two ingredients. It is easy to check that the composition law obtained in this way satisfies Eq.~\eqref{geodesic_tetrad}. This equation only determines the composition law for a given connection if one imposes the associativity property of the composition. Comparing with the previous reference, one then concludes that the composition law obtained from translations that leaves the form of the tetrad invariant is the associative composition law one finds by parallel transport, with the connection constructed from a tetrad and its derivatives as in Eq.\eqref{e-connection}. 

Finally, if the composition law is associative, then Eq.~\eqref{k-DCL} reduces to
\be
(p\oplus_k q)=p\oplus\hat{k}\oplus q.
\ee
Replacing $q$ by $(\hat{k}\oplus q)$ in Eq.~\eqref{geodesic_tetrad}, which is valid for any momenta ($p, q$), one obtains
\be
\frac{\partial^2  (p \oplus \hat{k} \oplus q)_\nu}{\partial p_\tau \partial(\hat{k} \oplus q)_\rho}+\Gamma^{\sigma \mu}_\nu (p \oplus \hat{k} \oplus q) \frac{\partial (p \oplus \hat{k} \oplus q)_\sigma}{\partial p_\tau}\frac{\partial (p \oplus \hat{k} \oplus q)_\mu}{\partial(\hat{k} \oplus q)_\rho}\,=\,0\,.
\ee
Multiplying by $\partial(\hat{k} \oplus q)_\rho/\partial q_\lambda$, one finds
\be
\frac{\partial^2  (p \oplus \hat{k} \oplus q)_\nu}{\partial p_\tau \partial q_\lambda}+\Gamma^{\sigma \mu}_\nu (p \oplus \hat{k} \oplus q) \frac{\partial (p \oplus \hat{k} \oplus q)_\sigma}{\partial p_\tau}\frac{\partial (p \oplus \hat{k} \oplus q)_\mu}{\partial q_\lambda}\,=\,0\,.
\label{connection_1}
\ee
Taking $p=q=k$ in Eq.~\eqref{connection_1}, one finally gets
\be
\Gamma^{\tau \lambda}_\nu (k)\,=\,-\left.\frac{\partial^2  (p\oplus_{k}q)_\nu}{\partial p_\tau \partial q_\lambda}\right\rvert_{p,q \rightarrow k}\,,
\ee
which is the same expression of Eq.~\eqref{k-connection} proposed in Ref.~\cite{AmelinoCamelia:2011bm}. This concludes that the connection of Eq.~\eqref{e-connection} constructed from the tetrad is the same connection given by the prescription developed in Ref.~\cite{AmelinoCamelia:2011bm} when the composition law is associative.

\section{Phenomenological consequences}
\label{sec:pheno}

Given the current stage of development of DSR, which we have described in the previous sections, phenomenological studies can rely on a framework to describe kinematics, while a full theory capable of describing the dynamical features of DSR is still missing. In this context, the  two main avenues of investigation concern particle propagation effect and effects due to the modified kinematics in interactions. 

Particle propagation effects have played a prominent role in the birth and the development of phenomenological studies in quantum gravity in general \cite{Amelino-Camelia:1997ieq, Ellis:2005sjy, Mattingly:2005re}, and are very relevant for the phenomenology of DSR models in particular \cite{Amelino-Camelia:2008aez, Addazi:2021xuf}. As we are going to discuss  more in detail in the following, the deformed kinematics described by DSR may lead to in-vacuo dispersion in particle propagation, producing shifts in the time of arrival of photons with different energies emitted simultaneously by the same source. Even though these effects will be in general suppressed by the ratio between the particle's energy and the Planck energy, they could be amplified significantly over large propagation distances, such as those characterizing astrophysical sources. In fact, sensitivity of astrophysical observations allows for meaningful constraints on Planck-scale suppressed time shifts, see the corresponding chapter in this book. 
Another class of propagation effects which may be present in DSR models is known as ``dual lensing'' \cite{Amelino-Camelia:2011ycm,Amelino-Camelia:2017pne, Amelino-Camelia:2016wpo}, and is such that the apparent direction from which astrophysical particles are emitted depends on their energy. This kind of effect has until now received less attention than the time shift, because current experiments have a limited source localisation capabilities. However,  future multi-satellite telescopes will significantly improve space-time localization of sources with respect to traditional  telescopes, possibly opening a window on dual lensing investigation.

Since the time shift effect is at the moment the one that is receiving the largest attention in theoretical and phenomenological studies, in the following we focus  on that.

\subsection{Propagation effects: time shift}

As we mentioned in the introduction, DSR models where originally conceived to provide a relativistic framework to encode 
Planck-scale modified energy-momentum dispersion relation which can induce potentially observable corrections to particles' speed of propagation. 

In the subsequent theoretical developments it was progressively understood that the emergence of the sort of effects that provided the original motivation is not a necessary consequence of DSR. Indeed, even when such time shift effects do emerge, they can take different quantitative dependence on the relevant quantities at play, including space-time curvature, which was ignored in the first studies. 

So while the original idea concerned quite a specific kind of propagation effect, 
these more recent findings provide us with a range of possibilities for effects we can search for in astrophysical data. On the one hand, it is important to be aware of all the possibilities, in order to make sure not to miss a discovery opportunity. On the other hand,  when some of these effects are excluded by experimental analyses, we get an essential guidance on the construction of a consistent DSR model that is compatible with observations. While one might worry that the variety of possibilities that at the moment seem to be compatible with the DSR framework might imply a lack of predictivity, one should consider that longitudinal propagation effects are not the only ones that can emerge in this framework (we already mentioned transverse propagation effects and effects on interactions), and that relativistic compatibility imposes compatibility conditions between the different effects.

The first DSR phenomenological studies were based on models such as the ones described in Section \ref{sec:dsr}, and in that context the velocity of particles was  simply deduced from the modified energy-momentum dispersion via $v={\partial E}/{\partial p}$. With this assumption, the first studies of the phenomenological consequence of Hopf-algebra deformations of relativistic symmetries \cite{Amelino-Camelia:1999jfz, Amelino-Camelia:2002siu} confirmed that the group velocity of plane waves defined in this framework would imply a momentum-dependent velocity for massless particles. This conclusion was challenged by studies claiming that alternative definitions of the velocity should be used \cite{Kosinski:2002gu,  Mignemi:2003ab, Daszkiewicz:2003yr,  Kowalski-Glikman:2003voa}, leading to standard propagation velocity.

The validity of  $v={\partial E}/{\partial p}$ ultimately relies on the assumption that a Hamiltonian description is still available, such that  $v \equiv {dx}/{dt} = \{x,H(p)\}$, where $\{...\}$ are Poisson brackets, and that phase space coordinates satisfy the usual relation $\{x, p\} = 1$, so that $x = {\partial}/{\partial p}$. 
For this reason, subsequent studies on the phenomenological consequences of DSR models \cite{Amelino-Camelia:2010wxx} relied on a covariant Hamiltonian formalism to derive the particle's worldline from 
\ba
\dot x^0&=\{x^0,C\}\,,\\
\dot x^i&=\{x^i,C\}\,,
\end{align}
where the role of the Hamiltonian is played by the (possibly deformed) Casimir of the DSR algebra (both in the Hopf algebra setting and in more general phenomenological models), e.g. given by \eqref{eq:kPCasimir}. The use of this formalism allows us to account for a possibly deformed symplectic structure of the phase space, namely a deformed bracket $\{x,p\}$ between coordinates and momenta. And in fact it turned out that, depending on the choice of this bracket (different choices being linked by momentum-dependent redefinitions of the coordinates), one could alternatively get standard \cite{Daszkiewicz:2003yr} or momentum dependent velocities for massless particles within the same momentum-space DSR description.

A solution to this puzzle was found \cite{Amelino-Camelia:2010wxx, Amelino-Camelia:2011ebd, Rosati:2012fb} when it was pointed out that it is not enough to look at the expression of the coordinate velocity to state whether a time shift effect is to be expected. % In fact, in presence of relativity of locality the equations of motion (leading to the worldlines) written by an observer are affected by coordinate artifacts when they are used to infer the behaviour of particles far away from the observer.
One needs to compare the observations made by observers local to the emission and to the detection of the particle whose propagation time is being computed. By doing this, one accounts for the possibly nontrivial action of the transformations linking the two observers.  In particular, for the model inspired by the $\kappa$-Poincar\'e algebra in the bicrossproduct basis (see Section \ref{sec:kPpheno}) it turned out that such nontrivial action compensates the effect of momentum-dependent redefinition of coordinates, so that the time shift effect cannot be reabsorbed in such a way \cite{Amelino-Camelia:2011ebd, Amelino-Camelia:2011uwb}.\footnote{
 One might reach different conclusions concerning time shifts when considering other DSR models, such as models based on different bases of the $\kappa$-Poincar\'e algebra. For example, using the  the classical basis of $\kappa$-Poincaré, there could be an absence of time shifts for massless particles with different energies~\cite{Carmona:2017oit}. Within the relative locality framework this can be understood in terms of the non-invariance of physical predictions under momentum space diffeomorphisms \cite{Amelino-Camelia:2019dfl}. 
 Moreover, depending on the effective scheme used for studying this effect,  different time delay formulas are obtained, and they may not lead to a time  shift~\cite{Carmona:2017oit,Carmona:2018xwm,Carmona:2019oph,Relancio:2020mpa}. }  This observation set the ground for the investigations that led to the relative locality proposal, discussed in Section \ref{sec:relloc}.

\begin{comment}

{\bf Exercise:} Following a similar procedure to that used to derive the worldlines \eqref{eq:rel_loc_worldline}, derive the worldlines obtained using spacetime coordinates $\chi^\mu$ such that 
\be
\{\chi^i,\chi^0\}=\frac{\chi^i}{\kappa}\,,\quad \{\chi^0,p_0\}=1\,,\quad \{\chi^i,p_j\}=\delta^i_j\,,\quad \{\chi^0,p_i\}=\frac{p_i}{\kappa}\,,\quad \{\chi^i,p_0\}=0\,.
\ee
To do so, notice that these coordinates are related by a momentum-dependent redefinition to those of eq. \eqref{eq:rel_loc_worldline} \cite{Amelino-Camelia:2013uya}, and that the free particle action needs to be transformed accordingly. Notice that from the worldlines one would deduce a momentum-independent velocity for massless particles.
\end{comment}

By performing the Hamiltonian analysis, and properly transforming from the reference frame of the emitter to that of the observer, one can show that for the $\kappa$-Poincar\'e model of Section \ref{sec:kPpheno} the expected time shift between a low-energy massless particle (not affected by DSR effects) and a high-energy massless particle (for which DSR effects are relevant) seen by the observer local to the detection is \cite{Amelino-Camelia:2011ebd, Amelino-Camelia:2011uwb, Amelino-Camelia:2013uya, Barcaroli:2015eqe}:
\be
\Delta t = - L \left(1-e^{-\frac{E}{\kappa}}\right)\simeq - L \frac{E}{\kappa}\,,\label{eq:timeshift}
\ee
where $L=T$ is the distance/time of flight between the source and the detector according to low energy particles\footnote{Remember that we set the low-energy particle velocity $c=1$.} and we also gave the first-order expression in powers of the particle energy over the energy deformation scale $\kappa$.   {\bf Exercise:} Derive Eq. \eqref{eq:timeshift}, following e.g. \cite{Amelino-Camelia:2013uya}. Notice that the first order expression in \eqref{eq:timeshift} coincides with the one that would be derived from the modified dispersion relation \eqref{eq:firstorderMDR} by using $v=\frac{\partial E}{\partial p}$ and $\frac{\eta}{E_P}=\frac{1}{\kappa}$.
 
In  recent studies on the relative locality framework, the importance of specifying the emission/detection mechanism has emerged \cite{Amelino-Camelia:2011uwb, Amelino-Camelia:2014qaa}. This is due to the fact that in order for the the action describing the interaction vertex \eqref{eq:rel_loc_action_total} to be covariant, one needs to consider the translation/boost generators associated to the total momentum of the vertex \cite{Amelino-Camelia:2011uwb, Gubitosi:2019ymi}, and this depends on the particles entering the interaction. Because of this, one may or may not predict a time shift effect, depending on the interactions at play \cite{Amelino-Camelia:2011uwb}. Whether this is a specific feature of the relative locality framework, due to the insistence on the use of a Lagrangian formalism, or whether a similar behaviour is to be expected in any DSR model where a Hamiltonian approach is used is still matter of investigation (see \cite{Gubitosi:2019ymi} for a discussion on the conceptual drawbacks of using a ``total momentum'' or ``total boost'' generator).

A somewhat parallel line of investigation has been focusing on how to include the effects of space-time curvature on the DSR-induced time shifts. Expressions such as \eqref{eq:timeshift} are valid for situations where space-time curvature can be neglected. However, given the cosmological distance of sources used to test this kind of effect, curvature should be taken into account, and it is generically expected to induce a dependence of the time shift on the redshift $z$ of the source. Until very recently, the great majority of studies assumed that one can simply replace the momenta appearing in the modified velocity of propagation with the physical momenta $p\rightarrow p/a$, where $a$ is the scale factor in a Friedmann-Robertson-Walker (FRW) metric. Then the time shift between a low-energy and a high-energy particle emitted simultaneously can be found by asking that they travelled the same comoving distance, and reads, for the $n=1$ case of \eqref{eq:MDR}:
\ba
\Delta t&=\eta \frac{ E_0}{E_P} D(z)\,, \label{eq:FRWtimeshift}\\ 
D(z)&=\frac{1}{H_0}\int_0^z d\zeta\frac{(1+\zeta)}{\sqrt{\Omega_m(1+\zeta)^3+\Omega_\Lambda}}\,,
\end{align}
with $H_0$, $\Omega_m$ and $\Omega_\Lambda$ denoting, respectively, the Hubble parameter, the matter fraction  and the cosmological constant in a FRW universe. $E_0$ is the energy of the high-energy particle measured today.
 {\bf Excercise:} compute the time shift \eqref{eq:FRWtimeshift} using the criteria described in the above paragraph. Use \cite{Jacob:2008bw} as guidance.
 
In  \cite{Rosati:2015pga, Amelino-Camelia:2020bvx} it was pointed out that one could in principle consider a more general dependence of the physical momenta on the scale factor. Moreover, in \cite{Rosati:2015pga} it was observed that the formula 
\eqref{eq:FRWtimeshift} is valid in the DSR scenario only when translational invariance is not deformed, otherwise one gets a more general dependence on the redshift of the source.\footnote{The result of \cite{Rosati:2015pga} was obtaining starting from a deformation of the relativistic transformations in de Sitter spacetime, which is maximally symmetric, and then deducing the time shift in FRW via a slicing procedure, first devised in  \cite{Marciano:2010gq}.}  In \cite{Barcaroli:2015eqe} the time shift expected in a de Sitter spacetime  with  Hopf-algebra deformation of the relativistic symmetries was computed.

\subsection{Modified interaction effects}

As we mentioned, the DSR framework is not currently embedded into a theory providing the dynamics of particles interactions. For this reason, we can only make arguments on the allowed interactions based on kinematical constraints.

Both modifications of the  dispersion relation and of the energy-momentum conservation law have a relevant role in this kind of analysis. As we discussed in Section \ref{sec:dsr}, relativistic consistency generally provides quite a rigid structure on the allowed  combinations of dispersion relation and conservation laws, and this affects crucially the kinematical analysis of interactions, making any possible modification very mild. 

In order to make this point clearer, let us start from an example where the relativistic consistency is violated, because there is a modified dispersion relation but everything else is the same as in special relativity. Clearly in this case the modified dispersion relation identifies a preferred frame, where it takes the specific form considered. This kind of scenario belongs to the framework of Lorentz Invariance Violation (LIV).

We are not going to review the LIV framework in detail, since this is the subject of another chapter in this book. Here we simply want to characterize the phenomenological differences between this and the DSR framework as far as interactions are concerned. Notice that since in both models one can envisage the emergence of modified dispersion relations, one expects in both cases propagation effects for particles traveling from astrophysical sources.\footnote{As we mentioned in the previous subsection, even  propagation effects might allow to distinguish between the two frameworks, since there could be a different dependence on the redshift of the source.} 

In the LIV framework, modified dispersion relations can have significant implications for certain decay processes \cite{Jacobson:2001tu, Amelino-Camelia:2001com}. For example, massless particles would be allowed to decay, so that  a process like  photon decay into an electron-positron pair ($\gamma \to e^+ + e^-$) would become possible. Using the dispersion relation \eqref{eq:MDR} with $n=1$, and assuming that the law of energy-momentum conservation is unmodified  with respect to the one of special relativity, as usually done in LIV studies, one  finds a relation between  the opening angle $\theta$ between the outgoing electron-positron pair and their energies $E_+, E_-$ [{\bf Exercise}]
\be
\cos\theta=\frac{2 E_+E_-+2 m_e^2 -2 \frac{\eta}{E_P} (E_+E_-^2+E_-E_+^2)}{2 E_+E_--m_e^2 \left(\frac{E_+}{E_-}+\frac{E_-}{E_+}\right)}\,.
\ee
For $\eta < 0$, the process is  always forbidden ($\cos\theta>1$), as in special relativity, but, for positive $\eta$ and $E_\gamma >> (m_e^2 E_P/|\eta|)^{1/3}$, one finds that one may have $\cos\theta < 1$. Notice that the energy scale $(m_e^2 E_P)^{1/3} \sim 10^{13}$ eV is within reach of astrophysics observations, and, in fact, strong constraints on $\eta$ have been set using the fact that we see photons of higher energies coming from astrophysical sources \cite{Jacobson:2001tu}.

Conversely, any DSR model must have stable massless particles. 
By establishing the existence of a threshold for photon decay one could therefore  falsify the DSR idea. In fact, an energy  threshold for the decay of massless particles cannot be introduced as an observer-independent law, and is therefore incompatible with the DSR principles.
Massless particles which are below a threshold energy value for one observer will be above that threshold for other boosted observers, and this would allow to identify a preferred frame. Let us see how this works in practice when considering the photon decay process discussed above when the kinematics is given by the DSR model considered in Section \ref{sec:dsr}, straightforwardly generalized to the 3+1 dimensions.

From the conservation of spatial momenta encoded in the addition law \eqref{eq:commutativeaddition} one finds that\footnote{We  write all formulas up to the first order in $\frac{\eta}{E_P}$. For ultra-relativistic electrons and positrons one can consider $m_e \frac{\eta}{E_P}\simeq 0$.}
\be
p_\gamma^2=p_+^2+p_-^2+2 p_+ p_-  \cos\theta+2 \frac{\eta}{E_P}( E_+ E_-^2+ E_- E_+^2)+2 \frac{\eta}{E_P}( E_+^2 E_- + E_-^2 E_+  )\cos\theta\,,
\ee
where $p_+,p_-$ and $p_\gamma$ are, respectively, the spatial momentum of the positron, the electron and the photon.
Using the dispersion relation \eqref{eq:firstorderMDR} this becomes
\begin{align}
E_\gamma^2+\frac{\eta}{E_P} E_\gamma^3=&E_+^2+E_-^2-2m_e^2+2 \left(E_+ E_--m_e^2\left(\frac{E_+}{E_-}+\frac{E_-}{E_+}\right) \right) \cos\theta+\frac{\eta}{E_P}(E_+^3+E_-^3) \nonumber\\
&+2 \frac{\eta}{E_P}( E_+ E_-^2+ E_- E_+^2)+3\frac{\eta}{E_P}\left( E_+ + E_-  \right)E_+ E_-  \cos\theta\,.
\end{align}
Using the conservation of energy encoded in \eqref{eq:commutativeaddition} one obtains:
\begin{align}
&E_+^2+E_-^2+2E_+E_- +\frac{\eta}{E_P} (E_++E_-)^3 +2 \frac{\eta}{E_P} (E_+ +E_-)E_+E_- \cos\theta=\nonumber\\ 
&E_+^2+E_-^2-2m_e^2+2 \left(E_+ E_--m_e^2\left(\frac{E_+}{E_-}+\frac{E_-}{E_+}\right) \right) \cos\theta +\frac{\eta}{E_P}(E_+^3+E_-^3) \nonumber\\
&
+2 \frac{\eta}{E_P}( E_+ E_-^2+ E_- E_+^2)+3\frac{\eta}{E_P}\left( E_+ + E_-  \right)E_+ E_-  \cos\theta\,.
\end{align}
From this, one finds that the correction terms cancel out and one recovers the standard special-relativistic result
\be
\cos\theta=\frac{2 E_+E_-+2 m_e^2}{2 E_+E_--m_e^2 \left(\frac{E_+}{E_-}+\frac{E_-}{E_+}\right)}\,,
\ee
that signals the impossibility of such decay (one always has $\cos\theta>1$).

While  massless particles decays are forbidden in DSR, one might still have modifications of the threshold for the decay of massive particles. However, it turns out that in this case  the modification of the thresholds  is only appreciable for energies of the order of the Planck scale $E_P$, which render them  unobservable in practice. Nevertheless, since in some DSR models there could be an absence of time delays~\cite{Carmona:2017oit,Carmona:2018xwm,Carmona:2019oph,Relancio:2020mpa}, one might assume that in these scenarios  the effective Planck energy is much lower, without being incompatible with current observations in particle accelerators~\cite{Albalate:2018kcf,Carmona:2021pxw} and in astroparticle physics~\cite{Carmona:2020whi,Carmona:2021lxr}

For similar reasons as the ones discussed above, in DSR  it is also forbidden to have a threshold below which a photon cannot produce electron-positron pairs in interactions with another  sufficiently high-energy photon. Such a process is indeed always allowed in special relativity, regardless of the energy of the low-energy photon, and this must be the case also in DSR scenarios. This is a necessary consequence of the fact that two relatively boosted observers attribute different energy to a given photon, and if there was a threshold then it would be possible to distinguish frames where the interaction can take place from those where the interaction cannot take place. 

{\bf Exercise:} Using again the 3+1 version of the model of Section \ref{sec:dsr} as done above, perform the kinematical analysis of the process $\gamma\gamma\to e^+ e^-$ focusing on a collinear process. Given some low energy $\epsilon$ for the low-energy photon, find the minimum energy $E_{min}$ of the high-energy photon for the process to take place. Show that within this model one recovers the same result as in special relativity, $E_{min}=\frac{m_e^2}{\epsilon}$. Use \cite{Amelino-Camelia:2011gae} as guidance.

\section*{Acknowledgements}
The authors would like to thank Claus Laemmerzahl and Christian Pfeifer for the invitation to contribute a chapter to this book  "Modified and Quantum Gravity - From theory to experimental searches on all scales - WEH 740", based on invited talks in the 740 WE-Heraeus-Seminar, 1-5 February 2021.

This work has been partially supported by Agencia Estatal de Investigaci\'on (Spain)  under grant  PID2019-106802GB-I00/AEI/10.13039/501100011033.
The authors also acknowledge participation in the COST Association Action CA18108 “Quantum Gravity Phenomenology in the Multimessenger Approach (QG-MM)”.

\bibliographystyle{h-physrev.bst}
\let\oldaddcontentsline\addcontentsline% Store \addcontentsline
\renewcommand{\addcontentsline}[3]{}% Make \addcontentsline a no-op
\bibliography{QuGraPhenoBib}
\let\addcontentsline\oldaddcontentsline% Restore \addcontentsline

\end{document}